\documentclass[twocolumn,showpacs,amsmath,amssymb,prx,superscriptaddress,longbibliography]{revtex4-2}

\usepackage[T1]{fontenc}
\usepackage{lmodern}
\usepackage{amssymb}
\usepackage{amsmath}
\usepackage{textgreek}
\usepackage{microtype}
\usepackage {longtable}
\usepackage{mathtools}
\usepackage{array}
\usepackage{bm}
\usepackage{tabularx}
\usepackage{booktabs}
\usepackage{makecell}
\usepackage{verbatim}
\renewcommand{\arraystretch}{1.2}

\usepackage{pgfplots}
\pgfplotsset{compat=1.18}
\usepackage{tikz}

\usepackage{float}

\usepackage{graphicx}
\usepackage{xcolor}
\usepackage{csquotes}
\usepackage{nccmath}
\usepackage[hyperindex,breaklinks]{hyperref}
\usepackage{makecell}

\newcommand \be{\begin{equation}}
\newcommand \ee{\end{equation}}
\newcommand \bea{\begin{eqnarray}}
\newcommand \eea{\end{eqnarray}}
\newcommand \bse{\begin{subequations}}
\newcommand \ese{\end{subequations}}
\newcolumntype{Y}{>{\raggedright\arraybackslash}X}

\begin{document}

%\preprint{}

\relpenalty=1
\title{Analytical emission model for the design of primary effusive sources}

\author{I.~N.~Ashkarin}
\affiliation {Universit\'e Paris-Saclay, CNRS, Laboratoire Aim\'e Cotton, 91405 Orsay, France}

\author{J.~Cheayto}
\affiliation {Universit\'e Paris-Saclay, CNRS, Laboratoire Aim\'e Cotton, 91405 Orsay, France}

\author{P.~Cheinet}
\affiliation {Universit\'e Paris-Saclay, CNRS, Laboratoire Aim\'e Cotton, 91405 Orsay, France}

\author{D.~Comparat}
\email{daniel.comparat@cnrs.fr}
\affiliation {Universit\'e Paris-Saclay, CNRS, Laboratoire Aim\'e Cotton, 91405 Orsay, France}

\author{S.~Lepoutre}
\email{steven.lepoutre@universite-paris-saclay.fr}
\affiliation {Universit\'e Paris-Saclay, CNRS, Laboratoire Aim\'e Cotton, 91405 Orsay, France}

\begin{abstract}
	
We present an analytical emission model that accurately predicts the properties of effusive sources formed by long collimation tubes. By construction, it captures the full range of molecular flow, from the transparent flux regime, which occurs in highly rarefied gases, to the opaque regime, which arises as the flux increases and interparticle collisions become non-negligible. The model is based on a previously developed secondary-emission-surface approach, improved here to overcome its internal limitations and recover the well-established axial flux intensity. It provides accurate analytical predictions of the angular intensity distribution in the molecular flow regime, offering valuable guidance for the design of efficient primary sources across a broad range of experiments in atomic and molecular physics.

\end{abstract}

\maketitle

\section{Introduction} \label{SecIntroduction}

Atomic and molecular beams are a central tool in modern physics and physical chemistry, with applications ranging from precision spectroscopy to surface science and fundamental studies of isolated collisions \cite{Ramsey1956,demarcus1956,1974mbld.rept....1A,GoodmanWachman1976,Bernstein1982,LevineBernstein1987,ScolesI,ScolesII,campargue2001atomic,Pauly,Lucas2013}. In many of these applications, a key requirement is the production of intense and well-collimated beams of neutral particles under vacuum. Designing such sources therefore requires reliable predictions of the emitted flux, angular distribution, and beam divergence, making modeling an essential step in the development of experimental setups \cite{gray1992design,ross95,Lucas2013,2013RScI...84j6113J,2015RScI...86b3105S,dotti2025robust}.
While numerical approaches such as Monte Carlo simulations \cite{nusinzon1977gas,shiwei2008computer,li2013modified,linke2025concurrent,kersevan2019recent} can in principle provide accurate descriptions, they are often too computationally demanding or insufficiently clear for rapid design optimization. This motivates the search for simple analytical models capable of capturing the essential physics while remaining practical for early-stage source design.

The theoretical description of beam emission is well established in the so-called molecular flow regime where  the particles are traveling collisionless before encountering any surface \cite{Ramsey1956,steckelmacher1986knudsen,ScolesI,ScolesII,laurendeau2005statistical,campargue2001atomic,pauly2013atom,Lucas2013}. In the simplest case of a thin-wall aperture, the emission is fully described by kinetic theory, leading to the well-known cosine (Lambertian) angular distribution for the beam emerging out of the oven. When a tube is added to improve collimation, the problem remains analytically tractable provided that the mean free path of the particles is much larger than the geometrical dimensions of the system. In this transparent (or Knudsen) regime, particles propagate ballistically between wall collisions, and the transmission and angular distribution can be described using Knudsen-Smoluchowski-Clausing theory and its extensions \cite{KnudsenAdP09I,KnudsenAdP09II,Smoluchowski1910kinetischen,steckelmacher1986knudsen,ClausingZfP30,ClausingAdP32,Lucas2013}. This framework provides a solid and widely used basis for the design of effusive sources operating at low pressure.

However, in order to achieve high flux, the source temperature must be increased, which raises the vapor pressure and reduces the mean free path. As a result, many practical sources operate in an intermediate regime where interparticle collisions inside the tube cannot be neglected. This so-called opaque molecular regime lies between the transparent Knudsen regime and more collisional regime, and leads to significant modifications of the density profile and emission properties of the beam. In this regime, the standard collisionless models fail to accurately predict key quantities such as the axial intensity, total flux, or angular divergence.

Several approaches have been proposed to extend  the  Knudsen-Smoluchowski-Clausing description  beyond the transparent molecular-flow regime,
including those of
Giordmaine and Wang \cite{GiordmaineJAP60}, Hanes \cite{HanesJAP60},
Becker \cite{Becker1961}, Ivanov and Troitskii \cite{IvanovTroitskii1963},
Zugenmaier \cite{ZugenmaierZaP66}, Olander \textit{et al.}
\cite{OlanderJAP69I,OlanderJAP69II,OlanderJAP70III,OlanderJAP70IV,OlanderJAP70V},
Beijerinck and Verster \cite{BeijerinckJAP75}, and Flory and Cutler
\cite{FloryJAP93}, to cite only a few (see also references in
\cite{steckelmacher1986knudsen,Lucas2013}).
These models incorporate, to varying degrees, effects such as density evolution along the tube or effective emission from secondary surfaces. Nevertheless, none of them provides a fully satisfactory compromise between physical accuracy, analytical simplicity, and ease of use for practical source design. This highlights the need for a simple and robust “toy model”  able to provide simple and reliable data  (let say at 10\% level) on the flow characteristics for both transparent and opaque regimes.

In this work, we propose a new analytical model of emission of an effusive source  valid for the overall molecular flow regime, that is from the transparent and the opaque regimes. Building upon the secondary-emission-surface approach introduced by Hanes \cite{HanesJAP60}, we reformulate the opaque source emission  as an equivalent transparent emission from an effective aperture. The position of which being chosen such as to reproduce the reference axial flux intensity derived by  Giordmaine and Wang \cite{GiordmaineJAP60}, leading to what we call the HGW (Hanes-Giordmaine-Wang)  model. This construction provides a simple and physically clear description that captures the main effects of collisional transport, while remaining analytically tractable. In addition to the axial intensity, the model yields accurate (at the required 10\% level) predictions for the total flux and angular width of the emitted beam, making it particularly suitable for early-stage design of effusive atomic sources.

The approach adopted in this article is deliberately pedagogical. In addition to introducing the model, we aim to clarify the hierarchy of physical assumptions, notations, and regimes that are often scattered across the literature and not always consistently defined. In Sec.~\ref{Foundations}, we define the problem, notations, and assumptions of the standard transparent regime. The approximations needed to deal with the opaque regime are introduced in Sec.\ref{SecToyModel}. Then a brief review of existing models for the opaque regime is given in Sec.~\ref{SecCompMod}. The proposed HGW analytical model is then presented in Sec.~\ref{SecAnMod}, followed by a discussion of its predictions and its relevance for practical source design.

\section{Theoretical Foundations: the transparent regime} 
\label{Foundations}

We begin by recalling the simplest and most fundamental configuration:
the emission of atoms through a thin-wall aperture separating a reservoir
at thermodynamic equilibrium from vacuum. Although this problem is treated
in many standard textbooks
\cite{Ramsey1956,ScolesI,ScolesII,campargue2001atomic,Pauly,Lucas2013},
we review it here for pedagogical and notational purposes. This reference
case provides the baseline against which all subsequent models are compared,
and it allows us to make explicit, step by step, the assumptions and
approximations that will later be used, modified or relaxed in the construction of
our model. 

Here and throughout, the term
``atom'' is used in a generic sense to denote either an atom or a
molecule. This simplification is justified because the present
description relies only on translational motion, wall interactions,
and interparticle collision mean free paths. Internal molecular
degrees of freedom (rotational or vibrational), chemical processes,
and state-changing collisions are not treated explicitly or will be mentioned if specifically required.

\subsection{Thin wall aperture reference case} \label{SubSecSourceDescription}

\begin{figure}[h]
	\includegraphics[width=0.99\columnwidth]{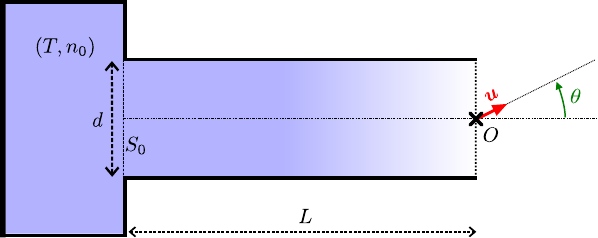}
	\caption{\label{FigOvenTube} Scheme of the example primary source considered for the presentation of emission models. The oven contains a gas at temperature $T$ and density $n_0$. The emitting nozzle is composed  of an aperture of diameter $d$ eventually followed by a single capillary tube of length $L$.}
\end{figure}

We first consider an atomic source, which comprises a reservoir containing vapor with particle number density $n_0$ and temperature $T$, assumed to remain under thermodynamic equilibrium conditions with pressure $P = n_0 k_B T$ (see Fig.~\ref{FigOvenTube}). In particular, the atomic velocities follow a Maxwell–Boltzmann distribution with mean thermal velocity
\begin{equation}
\bar v = \sqrt{\frac{8 k_B T}{\pi m}}.
\end{equation} where $m$ is the mass of the atom.

An aperture is then made to let atoms exit out of the oven. 
The properties of the emission are mainly governed by the value of the mean free path $\lambda$ inside the source gas, inversely proportional to its density, defined as:
\begin{equation}
\lambda = \frac{1}{\sqrt{2}\,\sigma n_0},
\end{equation}
where $\sigma = \pi d_{\rm kin}^2 $ is the collision cross-section with $d_{\rm kin}$ the so-called kinetic diameter.

In the so-called thin-wall aperture approximation, the aperture is assumed to be sufficiently thin that it does not behave as a channel. As a result, particles passing through the aperture experience no interparticle collisions within the aperture region. The reservoir is also assumed to remain in thermal equilibrium, so that the aperture does not disturb the Maxwellian velocity distribution inside the source. For this a necessary relevant condition is therefore
\begin{equation}
\lambda\gg d,
\end{equation}
where \(\lambda\) is the mean free path and \(d\) is the characteristic aperture size,
here taken as its diameter.
The precise boundary between flow regimes is
empirical and problem-dependent; however, values for the Knudsen number \(K \equiv \frac{\lambda}{d} > 10\) are commonly
taken as safe to enter fully the free-molecular, sometimes also named effusive, regime
\cite{karniadakis2005microflows,laurendeau2005statistical} (see also later defined Table \ref{tab:flow_regimes}).
Under these conditions, the flow through the aperture  and the thermodynamic properties of the vapor are well
described by the classical kinetic theory of dilute gases
\cite{Ramsey1956,ScolesI,ScolesII,campargue2001atomic,Pauly,Lucas2013}. 

In contrast, when \(\lambda\) becomes comparable to, or smaller than, \(d\),
interparticle collisions in the aperture region can no longer be neglected.
The flow then progressively leaves the effusive limit and enters collisional
regime
which are outside the scope of the present work.

The total number of
particles crossing the aperture per unit time $\dot N$ follows directly from kinetic theory and is obtained
by integrating the flux of particles moving toward the
opening. For the thin-wall (TW) aperature approximation, this yields the well-known result
\begin{equation}
\dot{N}_{TW} =  A \frac{ n_0 \bar v}{4} = \pi d^2 \frac{ n_0 \bar v}{16} 
\end{equation}
where $A=\pi d^2/4$ is the aperture area.
The factor \(1/4\) combines the restriction to velocities directed toward the aperture with the angular average of the normal velocity component relevant for the flux estimate.

The angular distribution of the emitted particles is
conveniently
described using  spherical coordinates with origin $O$ at the center of the nozzle and spherical coordinate  $0\leq \theta \leq \pi/2$ and $0 \leq \varphi < 2\pi$ with $ \theta$  the angle with respect to the normal to the
aperture
(see Fig.~\ref{FigOvenTube}). 
We define
the differential flux per unit solid angle per second 
$I(\theta,\varphi)$ as
the number of atoms flowing through the aperture within
 an infinitesimal  solid angle
$\sin(\theta) d\theta d\varphi$. For simplicity we deal in the whole article with cylindrical tubes that have rotational invariance 
so with $I(\theta) = I(\theta,\varphi)$, but the work can straightforwardly be adapted to non cylindrical (as rectangular for instance) tubes \cite{Pauly}.
The total flux is recovered by integrating over the forward
hemisphere
\begin{equation}
\dot N = \int_0^{2\pi} \!\!\!\!\!\!  d \varphi \int_0^{\pi/2} \!\!\!\!\!\!   I(\theta,\varphi)  \sin(\theta)  d \theta   =   2\pi \int_0^{\pi/2} \!\!\!\!\!\!   I(\theta) \sin(\theta)  d \theta  
\end{equation}
In principle
$I(\theta,\varphi)$  can be  ideally measured
 by considering a particle detector with an infinitesimal surface positioned in the far field, however in practice this measurement is not often easy to perform \cite{SteckelmacherJJAP74,krasuski1990gas}).

To characterize the angular intensity distribution, two quantities are commonly used: the axial intensity $I(0)$ and the normalized angular profile $f(\theta) $ linked by:
\begin{equation}
f(\theta) = \frac{I(\theta)}{I(0)}.
\end{equation}
In order to quantify the collimation of a source we can estimate the typical width of the angular profile by an angle  at half maximum $\theta_{1/2}$ by imposing 
\begin{equation}
f(\theta_{1/2})=1/2
\end{equation}

For the thin-wall  aperture emission case, the emission probability is proportional
to the normal component of the velocity. It is a
direct consequence of the isotropy of the velocity distribution
in the reservoir and constitutes a universal result
for effusive emission in the molecular-flow regime. This leads to 
the so-called cosine law (or Lambertian emission)  for the intensity $f_{TW}(\theta) = \cos\theta $.

We can thus summarize the  thin-wall (TW) aperture  results as:
\begin{align}
	I_{TW}(\theta) &= I_{TW}(0) \cos\theta \label{EqITW} \\
    I_{TW}(0) &= \frac{n_0 \bar v d^2}{16} \label{EqI0Tr} \\
    \dot{N}_{TW} &= \pi  \frac{n_0 \bar v d^2}{16} = \pi I_{TW}(0)  \\
    {\theta_{1/2}}_{TW} & = \pi/3
\end{align}

The cosine angular distribution implies
that most atoms are emitted at large angles
(${\theta_{1/2}}_{TW}$ is 60 degree)
resulting in a
poorly collimated beam. Any improvement in collimation
must therefore rely on additional geometrical constraints,
such as the introduction of a (or multiple) tube to reach
a ``well-collimated'' source (or beam) that should ideally satisfy $\theta_{1/2} \ll 1$ rad.

\subsection{Addition of a tube: different flow regimes}
\label{sec:hypotheses}

In order to improve the collimation of the atomic flux, an additional capillary tube of length $L$ is added to the source output (see Fig.~\ref{FigOvenTube}). This introduces a second characteristic length scale $L$ in addition to the aperture diameter $d$. It is convenient to define the aspect ratio as
\begin{equation}
    \Gamma = \frac{L}{d}
\end{equation}
The presence of the tube modifies the emission in two ways: it reduces the total transmitted flux, since only a fraction of particles entering the tube reaches the exit, and it affects particle trajectories, thereby narrowing the angular distribution. The price paid for collimation is thus a finite geometrical impedance of the tube.
In this article we will mainly consider
 well-collimated sources, corresponding to long tubes with $\Gamma \gg 1$. 

The properties of the flow inside the tube are governed by the mean free path $\lambda$, which sets the importance of interparticle collisions
depending on the  Knudsen number $K_L = \frac{\lambda}{L}$.
 Following \cite{OlanderJAP70III, BeijerinckJAP75}, we prefer to define its inverse which is proportional to the density $n_0$ inside the oven and can thus be seen as a reduced oven density (or opacity) by:
\begin{equation}
    n_0^* = \frac{L}{\lambda} 
\end{equation}
This dimensionless parameter compares the tube length to the mean free path. It can be intuitively linked to the mean number of collisions arising along the tube. It  allows to distinguish two new different transport regimes within the molecular regime itself: the transparent regime and the
opaque molecular regime summarized in Table \ref{tab:flow_regimes}.

\begin{table}[h]
\centering
\begin{tabular}{|l|l|l|c|c|}
\hline
Regime & condition & Subregime & \multicolumn{2}{c|}{Sub-condition} \\
\hline
\makecell[l]{Molecular,\\ effusive flow}
& \(\lambda > d\)
& Transparent
& \(\lambda > L\)
& \(n_0^* < 1\) \\
\cline{3-5}
&
& Opaque
& \(\lambda < L\)
& \(1 < n_0^* < \Gamma\) \\
\hline
\makecell[l]{Collisional\\flow}
& \(\lambda < d\)
& --
& --
& \(n_0^* > \Gamma\) \\
\hline
\multicolumn{5}{|l|}{\textit{Walls:} bright-wall (re-emitting) / dark-wall (sticking)} \\
\hline
\end{tabular}
\caption{
Indicative flow-regime and tube-wall classification  for a (long) tube of diameter \(d\), length
\(L\), according to the value of the  mean free path \(\lambda\) or of the dimensionless  (length ratio) parameters:
aspect ratio \(\Gamma=\frac{L}{d} \) and $n_0^* = \frac{L}{\lambda}$.
The inequalities only indicate nominal
boundaries. In practice, transitions are smooth (see Fig. \ref{FigAngWidthOp} for instance); thus safety factors of order \(10\) are often used to define
well-separated regimes \cite{laurendeau2005statistical}.
 The terminology used in
the table is the one adopted throughout this work: molecular (or effusive) flow  is divided
into transparent and opaque molecular regimes, followed by a collisional
regime. Other terms are sometimes used in the literature but are not always fully standardized: the
transparent molecular regime is also often called free-molecular,
collisionless, effusive, or Knudsen-limit flow; the opaque molecular regime is
also referred to as intermediate, transition(al), collision-affected molecular
flow, Knudsen-diffusive flow, or diffusive molecular flow; and the collisional
regime is commonly called continuum, viscous, hydrodynamic, or gas-dynamic
flow, with the slip-flow regime sometimes used for the near-continuum
transition. The bright-wall/dark-wall
terminology refers only to wall interaction boundary conditions and is
independent of the transport-regime classification.
}
\label{tab:flow_regimes}
\end{table}

The limit \(n_0^* \ll 1\) (i.e. \(\lambda \gg L\)) corresponds to the
\textit{transparent} regime, by analogy with optics: interparticle collisions
inside the tube are negligible. Particles then propagate ballistically between
wall collisions, and the emission properties depend only on the tube geometry,
through its aspect ratio \(\Gamma\).
When the density is increased so that \(n_0^*\) becomes of order unity or
larger, while remaining below \(\Gamma\) (i.e. \(d \ll \lambda \lesssim L\)),
the system enters the \textit{opaque} molecular regime. In this regime,
interparticle collisions occur inside the tube and progressively modify the
density profile and angular distribution, while the mean free path remains
larger than the tube diameter.
Finally, when \(n_0^*\) becomes of order \(\Gamma\) or larger
(i.e. \(\lambda \lesssim d\)), the system leaves the molecular-flow regime and
enters a collisional regime, which will not be discussed here.

\subsection{Re-emission hypothesis and Clausing formulation for the flux}
\label{sec:hypotheses}

We first restrict ourselves to the transparent regime, where only particle-tube wall collisions need to be considered. The opaque regime will follow by adding particle-particle collisions.

The theoretical description of transport in the transparent regime relies on
assumptions regarding atom--wall interactions. Depending on the chemical nature
of the species and the wall material, two limiting behaviors can be distinguished: Dark-wall regime when particles stick to the wall and bright-wall regime when  particles are nearly instantaneously re-emitted.

In the first case, collisions lead to sticking of the particles on the wall,
possibly followed by dimer, cluster, or film formation; this defines the
so-called dark-wall regime \cite{drullinger1991nist,de2021spider,HahnRSI22}.
The dark-wall regime is  simple to describe, as it effectively
reduces to an angular filtering of the thin-wall aperture emission, it narrows
the angular distribution while strongly reducing the transmitted flux.
Indeed, in transparent regime only the direct trajectories passing through the aperture output matters and the dark-wall is thus identical to a two aperture setup.
This
behavior is often difficult to avoid for highly
reactive species such as alkali atoms (e.g.\ cesium)
\cite{pan2013modeling,HahnRSI22}. However, the choice of a dark-wall tube can be advantageous for beam collimation
\cite{li2019cascaded,zhang2025foam}.

In the present work, we restrict ourselves to the opposite, bright-wall limit,
which is relevant for high-brightness sources. In this limit, particles impinging
on the tube walls are rapidly reemitted, with negligible residence time. The central
assumption is the Knudsen law for gas--surface scattering: each surface element
reemits incident particles diffusely, with a cosine angular distribution relative to
the local normal and independently of the incoming direction \cite{KnudsenAdP09I, KnudsenAdP09II, ClausingAdP32,steckelmacher1986knudsen}.  
The validity of this description requires that
additional processes -- such as adsorption, surface diffusion, chemical reactions,
inelastic effects, partial specular reflection, deviations from an ideal cosine law or interparticle collisions -— do
not significantly alter the transport. These effects can become important in
practice, particularly for reactive species such as alkali atoms
\cite{Drullinger85, DaviesJPD83,king1975thermal,matsushima2003angle,LevdanskyIJHMT08,Lucas2013, HahnRSI22,BernasekPSS75,MurphyJVST89, EiblJVST98,SchwarzSelingerJVST00,LibudaRSI00,HurlbutRRMB59,LoganJCP66,EdmondsJVST65,HobsonJVST69,DavisJAP64, AquilantiJCP99}. 

Under these assumptions, each wall collision completely erases the memory of
the incoming direction, so that molecular flow through the tube consists of
ballistic flights at the mean thermal speed \(\bar v\), interrupted by diffuse,
cosine-law wall reemission. The speed \(\bar v\) is evaluated at the reservoir
temperature \(T\), assuming an isothermal tube, although slightly higher wall
temperatures are often used in practice to prevent condensation and clogging
\cite{ross95}. In long tubes, repeated wall collisions progressively randomize the
trajectories, giving a Knudsen-type transport that may be viewed macroscopically
as diffusion-like. Note that the term ``diffuse'' is used here in two distinct
senses: locally, it refers to the cosine-law angular distribution of wall
reemission, while macroscopically it describes the resulting diffusion-like
transport. 

Clausing reformulated the tube-transport problem in terms of a
transmission probability, or Clausing factor, \(W(\Gamma)\), which is the
fraction of particles entering the tube that finally  exit into vacuum
\cite{ClausingZfP30,ClausingAdP32}. In the transparent Knudsen regime,
where interparticle collisions are negligible, this factor depends only on
the tube geometry and provides a compact measure of its impedance relative
to a thin-wall aperture. The total transmitted flux is then
\begin{equation}
    \dot{N}
    =
    \dot{N}_{\mathrm{TW}}\,W(\Gamma),
    \label{EqTotFluxGeneral}
\end{equation}
where \(\dot{N}_{\mathrm{TW}}\) is the thin-wall aperture flux for the same
entrance area.

It is important to stress that this regime is not a local-equilibrium regime.
Particles propagate ballistically between wall collisions, and the cosine law
enters only as a boundary condition for diffuse wall reemission. Clausing's
approach therefore solves a non-equilibrium geometrical transport problem:
successive wall collisions are accounted for through an integral equation
over particle trajectories, without requiring a local temperature, an
isotropic velocity distribution, or an assumed density profile inside the
tube.
The exact value of \(W(\Gamma)\) follows from Clausing's original integral
treatment, or equivalently from later numerical and empirical evaluations
\cite{Lucas2013,Pauly}.
  A convenient interpolation is given by
\begin{equation}
    W_{Clau}(\Gamma)
    =
    \frac{\frac{4}{3\Gamma}}
    {1+\frac{4}{3\Gamma}} 
    \label{EqWClau}
\end{equation}

It has the proper limits
$W \xrightarrow[\Gamma \to 0]{} 1,\;
W \xrightarrow[\Gamma \to \infty]{} \frac{4}{3\Gamma} $ and furthermore is  valid (absolute accuracy of a few $10^{-3}$ \cite{Lucas2013}) over all $\Gamma$ values.
Unless specified otherwise, this expression will be used in the following as a practical approximation of the tube impedance.

We can compare it to the asymptotic dark-wall result where  basic geometrical consideration \cite{HahnRSI22} lead to a transmission probability
\( W_{DW}=  1/\Gamma^2 \). We thus see the  that a dark-wall is more than $ \Gamma $ less transmissive than a bright-wall case. This  is thus a huge loss factor for large $ \Gamma $ and explain why we deal here  only with the bright-wall case.

\section{Linear-density approximation for collisional tube emission}
\label{SecToyModel}

\subsection{Density profile and physical assumptions}

When increasing the source temperature, interparticle collisions inside the tube can no longer be neglected. This regime is less amenable to simple analytical treatment, and discrepancies between predictions based on collisionless models and experimental observations are commonly reported \cite{GiordmaineJAP60,HanesJAP60,AdamsonVacuum88II}. 
Indeed,
the breakdown of the transparent regime has several important consequences:
the density is modified by collisional transport and is no longer purely geometrical,
 particles may be redirected toward the tube exit through collisions enhancing axial intensity,
and collisions degrade collimation by redistributing velocities.

To go further and to be able to calculate  intensity distribution, we thus need to develop
alternative approaches introducing additional levels of approximation. The mostly used one is by reinterpreting the transport in terms of a local density profile $n(z)$ assuming that particles can be described locally through a scalar density \cite{Pauly,Lucas2013}. A second hypothesis is  that wall re-emission can still be treated effectively as a cosine angular distribution at each position along the tube. This allows one to reconstruct angular distributions and fluxes from $n(z)$, at the price of implicitly reintroducing a form of local equilibrium or isotropy. 
A major advantage of the local-density approach is that it becomes
physically better justified as the system enters the opaque regime: increasing
interparticle collisions progressively restore local equilibration inside the tube,
making a local thermodynamic description more meaningful.

A first simplifying assumption is transverse invariance, so that the density
depends only on the longitudinal coordinate \(z\). This quasi-one-dimensional
description is expected to hold for long tubes, \(\Gamma \gg 1\), sufficiently far
from the entrance and exit regions. Numerical studies of molecular flow support
this approximation, showing that radial density variations remain within our
\(10\%\) accuracy budget in this limit even if larger
deviations may occur for short tubes \cite{DaviesJPD83,FloryJAP93,shiwei2008computer,he2025spatial}.
This is also the reason why we restrict the following discussion to long tubes,
typically \(\Gamma>10\), for which also entrance, bulk, and exit regions can be
meaningfully separated.

Under these conditions, the longitudinal density profile can be usefully
approximated by a linear function \cite{Lucas2013}. 
Following the notation of \cite{OlanderJAP70III}, we write
\begin{equation}
    n(z) = n_0 \left(\zeta_1 + \frac{\zeta_0 - \zeta_1}{L} z \right),
	\label{EqScDensityVsZOp}
\end{equation}
where $n_0$ is the reservoir density and $\zeta_1$ corresponds to the upstream ($z=0$ entrance)  and $\zeta_0$ to the downstream ($z=L$ exit) dimensionless coefficients characterizing the boundary conditions at the tube ends (so-called end effects). The linear approximation should
be understood as an effective description of the progressive particle loss along
the tube, rather than as a thermodynamic boundary condition imposed at its ends.
 These coefficients account for the fact that
the naive expression $n(z) = n_0(1-z/L)$ might be too crude (remark that it explains the notation 0 and 1 because here \(\zeta_0 =0\) and \( \zeta_1 =1 \)) because the presence of the tube modifies the local density near its boundaries compared to the ideal reservoir or vacuum limits.  In particular, even in the transparent regime, the density at the tube entrance is generally reduced compared to $n_0$, while the density at the exit does not strictly vanish and is not a true equilibrium density, since only forward velocities are populated. As a result, the effective density associated with the outgoing flux scales as $n(L) \sim n_0 W/2 $, so $\zeta_0 \sim W/2 $, reflecting both the transmission probability $W$ and the restriction to half of velocity space.

The linear profile of Eq.~\eqref{EqScDensityVsZOp} can be interpreted as the simplest interpolation between  two boundary conditions. It is not an exact solution of the transport problem, but it captures the dominant behavior of the density in long tubes. This approximation has been shown to provide a good description of molecular flow both in the transparent regime \cite{HelmerJVST67II,BeijerinckPhysica76} and, to a certain extent, in the opaque regime \cite{KurepaJAP81, deLeeuwUTN66}, which justifies its use as a basis for simple analytical models.

An important feature of this formulation is that all geometrical and transport effects are now encapsulated in the quantities $\Gamma$, $\zeta_0$, $\zeta_1$ and $n_0^*$. In particular, once $n(z)$ is specified, the evaluation of the emitted intensity reduces to a purely geometrical problem involving the propagation of particles from different positions inside the tube to the far field.

\subsection{Angular profiles}
\label{SecAngularProfiles}

We now follow the model and notations of \cite{OlanderJAP70III}. The main advantage is that it encounters all other model we are going to mention. Because some typos exists in the literature   \cite{Note1,Note2} we  think it is worth to recall all formulas here, even if they are sometimes cumbersome. 

A key feature of the Olander and Kruger \cite{OlanderJAP70III}  approach is that it extends the geometrical
description of tube emission beyond the strictly collision-free limit, while
remaining within the molecular-flow regime. It is intended for the range in
which interparticle collisions modify the angular distribution and the
centerline intensity, but do not yet significantly change the total Clausing
flow; hydrodynamic effects are therefore excluded but the opaque regime can be treated.

The calculation proceeds by prescribing the  linear
longitudinal density profile
\(n(z)\) and using it to estimate local gas-phase
and wall-collision rates. The outgoing angular distribution is then constructed
as the sum of three contributions: particles transmitted directly from the
reservoir, particles diffusely reemitted by the walls, and particles scattered
toward the exit after interparticle collisions inside the tube (the relative importance of each term is discussed in Ref. \cite{FloryJAP93}). In this sense,
collisions enter statistically, through attenuation and redistribution terms,
rather than through explicit tracking of the full phase-space distribution.

This closure is approximate and the
relations between density and collision rates used in such calculations are
equilibrium-gas relations, whereas the gas in the tube is generally neither
Maxwellian nor angularly isotropic. The model therefore replaces the full
Boltzmann transport problem by an effective one-dimensional description based
on \(n(z)\), diffuse wall reemission, and local statistical treatment of
interparticle collisions. It neglects velocity-space anisotropies, correlations
between successive collisions, and hydrodynamic effects.

\paragraph{Geometrical quantities.}

We  first introduce three dimensionless quantities that encode the geometry of the tube:
\begin{equation}
\begin{split}
    \theta_o &= \arctan \frac{1}{\Gamma}, \\
    q &= \Gamma \tan \theta, \\
    R(q) &\underset{q \leq 1}{=} \arccos q - q\sqrt{1-q^2}.
\end{split}
\label{EqDefTheta0QR}
\end{equation}

$\theta_o$  corresponds to the limiting angle above which the tube walls completely mask the direct line-of-sight from the oven to the observer. The parameter $q$ is a convenient rescaled angular variable incorporating the tube geometry. For $\theta \leq \theta_o$ (i.e. $q \leq 1$), the function $R(q)$ describes the geometrical reduction of the visible aperture due to partial masking by the tube. Indeed  in the ideal (\( \zeta_0=0, \zeta_1=1\) ) transparent regime, from the first aperture the flux $I_0(\theta)$ that goes out  is  \cite{Pauly}
 \(I_0(\theta) =  I_{TW}(\theta) \frac{2}{\pi} R(q) \). 

\paragraph{Density and collisional parameters.}

The following quantities incorporate the effects of the density profile and, more generally, of collisional processes:
\begin{equation}
\begin{split}
    \delta_0 &= \sqrt{\frac{n_0^*}{2}} \frac{\zeta_0}{\sqrt{\zeta_1 - \zeta_0}},  \\
    \delta_1 &= \sqrt{\frac{n_0^*}{2}} \frac{\zeta_1}{\sqrt{\zeta_1 - \zeta_0}} 
\end{split}
\label{EqDefDelta01}
\end{equation}

They arise from the integration of the linear density profile and characterize the strength of collisional redistribution along the tube. In the transparent limit ($n_0^* \to 0$), these quantities vanish and the expressions simplify considerably.

We also define the auxiliary function

\begin{equation}
\begin{array}{l}
S(q) = \int_0^q dz\, \sqrt{1-z^2} \\
\quad \times \Bigg[
\operatorname{erf}\!\left(
\frac{\delta_0}{\sqrt{\cos \theta}}
+ z \sqrt{\frac{n_0^*}{2}}
\frac{\sqrt{(\zeta_1-\zeta_0)\cos\theta}}{\Gamma\sin\theta}
\right)
- \operatorname{erf}\!\left(
\frac{\delta_0}{\sqrt{\cos \theta}}
\right)
\Bigg] \nonumber
\end{array}
\label{EqDefSofQ}
\end{equation}

which accounts for the cumulative effect of collisions along trajectories inside the tube. Although the definition appears singular for $\theta=0$, the function can be extended by continuity, with $S(0)=0$.

\paragraph{Axial intensity normalization.}

The angular profiles are normalized using the reduced axial intensity
\begin{equation}
\begin{split}
    A(n_0^*,\zeta_0,\zeta_1) 
    &= \frac{I(0)}{I_{TW}(0)} \\
    &= \frac{\sqrt{\pi}}{2} \sqrt{\frac{2}{n_0^*}} 
    \exp(\delta_0^2)\sqrt{\zeta_1-\zeta_0} \\
    &\quad \times \left[\text{erf}(\delta_1) - \text{erf}(\delta_0)\right] \\
    &\quad + \zeta_0 + (1-\zeta_1)\exp(\delta_0^2-\delta_1^2).
\end{split}
\label{EqRedAxialIntensityOp}
\end{equation}
The chosen notation $A$ emphasizes both its axial character (on-axis intensity) and its role as an attenuation (or transmission) factor relative to the thin-wall reference.

As will be discussed later, this quantity depends only weakly on the precise values of the end effects, and is primarily governed by the reduced density $n_0^*$. In practice, it can often be approximated as a function $A(n_0^*)$.

\paragraph{General angular profiles.}

 The angular profiles can be expressed as

\begin{equation}
\begin{split}
    &f(\theta,n_0^*,\Gamma,\zeta_0,\zeta_1)
    \underset{\theta \leq \theta_o}{=} \frac{1}{A(n_0^*,\zeta_0,\zeta_1)} \Biggl(
    \zeta_0 \cos \theta \\
    &\quad + \frac{2}{\pi}\cos\theta \Biggl\{
    \frac{\sqrt{\pi}}{2} \sqrt{\frac{2}{n_0^*}} 
    \exp\left( \frac{\delta_0^2}{\cos \theta}\right) \sqrt{(\zeta_1-\zeta_0)\cos\theta} \\
    &\qquad \times \left[
    \left(\text{erf}\left(\frac{\delta_1}{\sqrt{\cos \theta}}\right) - \text{erf}\left(\frac{\delta_0}{\sqrt{\cos \theta}}\right)\right) R(q) 
    + 2 S(q)
    \right] \\
    &\qquad + (1-\zeta_1)\exp\left( \frac{\delta_0^2}{\cos \theta} - \frac{\delta_1^2}{\cos \theta}\right) R(q)
    \Biggr\}
    \Biggr),
\end{split}
\label{EqProfAngOpLowAngle}
\end{equation}

\begin{equation}
\begin{split}
    &f(\theta,n_0^*,\Gamma,\zeta_0,\zeta_1)
    \underset{\theta \geq \theta_o}{=} \frac{1}{A(n_0^*,\zeta_0,\zeta_1)} \Biggl(
    \zeta_0 \cos \theta \\
    &\quad + \frac{2}{\sqrt{\pi}} \sqrt{\frac{2}{n_0^*}} 
    (\cos\theta)^{3/2} \exp\left( \frac{\delta_0^2}{\cos \theta}\right) \sqrt{\zeta_1-\zeta_0}\, S(1)
    \Biggr).
\end{split}
\label{EqProfAngOpHighAngle}
\end{equation}

These expressions naturally separate the contributions of (i) direct emission from the tube entrance, (ii) attenuation due to geometrical masking, (iii) re-emission from the tube walls, and (iv) collisional redistribution inside the tube.

\paragraph{Transparent limit.}

In the transparent regime ($n_0^* \to 0$), collisional effects vanish and the expressions simplify considerably. One obtains

\begin{align}
    f_{tr}(\theta,\Gamma,\zeta_0,\zeta_1)
    &\underset{\theta \leq \theta_o}{=} 
    \zeta_0 \cos \theta \nonumber \\
    &\quad + \frac{2}{\pi}\cos\theta \Biggl[
    (1-\zeta_0) R(q) \nonumber \\
    &\qquad + \frac{2}{3q}(\zeta_1-\zeta_0)
    \left(1-(1-q^2)^{3/2}\right)
    \Biggr],
\label{EqProfAngTrLowAngle}
\\
    f_{tr}(\theta,\Gamma,\zeta_0,\zeta_1)
    &\underset{\theta \geq \theta_o}{=} 
    \zeta_0 \cos \theta 
    + \frac{4}{3\pi q} (\zeta_1-\zeta_0)\cos\theta.
\label{EqProfAngTrHighAngle}
\end{align}

\section{Emission models}
\label{SecCompMod}

In this section, we review some key models proposed in the literature for molecular-flow emission through tubes. These models can be interpreted as different prescriptions for $\zeta_0$ and $\zeta_1$, reflecting distinct physical assumptions about the transport inside the tube.

\subsection{Computational models}
\label{sec:computationalmodel}

We now describe several models that attempt to capture emission in different regimes. We refer to them as \textit{computational models}, as they rely on explicit evaluation of the previous expressions using prescribed values of $\zeta_0$ and $\zeta_1$.

\paragraph{Clausing model (transparent regime).}

The \textit{Clausing emission model} \cite{ClausingZfP30,ClausingAdP32} applies strictly to the transparent regime. In the long-tube limit, it prescribes
\begin{equation}
    \alpha = \frac{2}{3\Gamma},  \qquad    \zeta_0 = \alpha \approx    \frac{W_{Clau}(\Gamma)}{2}, \qquad \zeta_1 = 1-\alpha, 
\end{equation}

\begin{figure}[h]
    \centering
    \begin{tikzpicture}
    \begin{axis}[
        width=0.99\columnwidth,
        height=0.62\columnwidth,
        xlabel={$\Theta$},
        ylabel={$f_{\mathrm{Clau}}(\Theta,\Gamma)$},
        xmin=0,
        xmax=1,
        ymin=0,
        xlabel style={
            at={(axis description cs:0.5,0)}, % move UP (closer to axis)
            anchor=north
        },
        ylabel style={
            at={(axis description cs:0.05,1)}, % move RIGHT (closer)
             anchor=south,
             rotate=270
        },
        grid=major,
        line width=1pt,
        tick style={black},
        label style={font=\small},
        tick label style={font=\small},
        legend style={
            font=\small,
            draw=none,
            at={(0.97,0.97)},
            anchor=north east
        },
    ]

    \addplot[thick, red]
        table[
            col sep=comma,
            ignore chars={"},
            trim cells=true,
            x index=0,
            y index=1
        ] {Clausing_LT.csv};

    \addplot[thick, green!60!black]
        table[
            col sep=comma,
            ignore chars={"},
            trim cells=true,
            x index=0,
            y index=2
        ] {Clausing_LT.csv};

    \addplot[thick, blue]
        table[
            col sep=comma,
            ignore chars={"},
            trim cells=true,
            x index=0,
            y index=3
        ] {Clausing_LT.csv};

    \legend{$\Gamma=10$, $\Gamma=25$, $\Gamma=100$}

    \end{axis}
    \end{tikzpicture}
    \caption{\label{FigProfilesTrVariousGamma}
    Examples of Clausing angular profiles in the transparent regime for selected values of the aspect ratio $\Gamma$ within the long-tube limit.}
\end{figure}
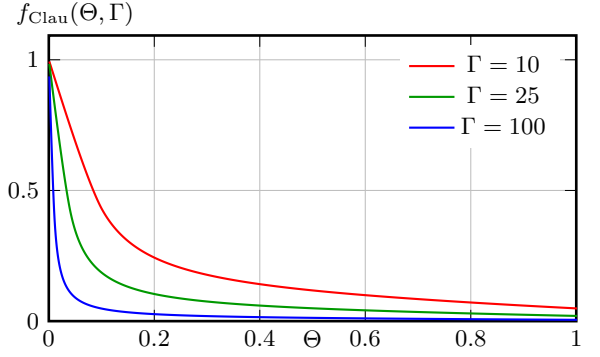

The corresponding angular profiles are obtained from
\begin{equation}
    f_{Clau}(\theta,\Gamma) = f_{tr}(\theta,\Gamma,\alpha,1-\alpha)
    \label{EqfClau}
\end{equation}
and examples are shown in Fig. \ref{FigProfilesTrVariousGamma}.
This model provides an exact reference for collisionless transport and correctly captures the purely geometrical collimation of long tubes.

\paragraph{Giordmaine--Wang model.}

The model proposed in Ref.~\onlinecite{GiordmaineJAP60} 
represents an early and influential
attempt to bridge the transparent  and the opaque molecular regime.
It assumes
\begin{equation}
    \zeta_0 = 0, \qquad \zeta_1 = 1,
\end{equation}
thereby neglecting end effects entirely.

This assumption greatly simplifies the analytical treatment, but introduces significant limitations. In particular, wall emission is underestimated, leading to a strong suppression of large-angle contributions and an underestimation of the total flux (typically by $\sim 50\%$).

Despite these limitations, the model provides two important results.

First, an analytical expression for the emission width in the transparent limit:
\begin{equation}
    \theta_{1/2,tr} =  \theta_{1/2,GW}  \simeq 1.68 \frac{a}{L} = \frac{0.84}{\Gamma}.
    \label{EqThetaHWTr}
\end{equation}
indicating clearly the collimation gain when using a tube to improve beam divergence. 
Interestingly we can compare it to the dark-wall result \cite{HahnRSI22}   \( \theta_{1/2,tr,DW} = \frac{0.5}{\Gamma} \) indicating that the bright-wall does no significantly increase the beam divergence compare to the pure filtering effect of a dark-wall.

Second, a remarkably accurate expression for the axial intensity, which is largely insensitive to end effects:
\begin{equation}
    A_{\rm GW}(n_0^*) = \frac{\sqrt{\pi}}{2} \sqrt{\frac{2}{n_0^*}} 
    \,\text{erf}\!\left(\sqrt{\frac{n_0^*}{2}}\right).
    \label{EqDefAGiordmaine}
\end{equation}

This result has been confirmed by later studies \cite{OlanderJAP69I, BeijerinckJAP75} and constitutes a key reference for modeling axial emission.

\paragraph{Zugenmaier model (reference model).}

The \textit{Zugenmaier model} \cite{ZugenmaierZaP66,Note3}) is generally regarded as one of the most accurate descriptions of molecular-flow emission \cite{LucasVacuum73,Lucas2013,he2025spatial}. It combines a refined estimate of the Clausing factor with a consistent treatment of the density profile.

The transmission probability is approximated by
\begin{equation}
    W_Z(\Gamma) = \frac{4\Gamma^3 + 6\Gamma +4 - 4(\Gamma^2+1)^{3/2}}{2\Gamma^3 + 6\Gamma +2 - 2(\Gamma^2+1)^{3/2}},
    \label{EqWZugenmaier}
\end{equation}
and the corresponding boundary conditions are
\begin{equation}
    \zeta_{0,Z} = \frac{W_Z(\Gamma)}{2}, \qquad \zeta_{1,Z} = 1.
\end{equation}

The angular profiles are then given by
\begin{equation}
    f_Z(\theta,n_0^*,\Gamma) = f\left(\theta;n_0^*,\Gamma,\frac{W_Z(\Gamma)}{2},1\right).
    \label{EqJZugenmaier}
\end{equation}

This model provides accurate predictions for both axial intensity and angular distribution over a wide range of conditions. However, it relies on relatively complex expressions and is not easily amenable to simple analytical interpretation.

\paragraph{Phenomenological extensions (Lucas).}

Due to the complexity of the full expressions, Lucas performed a detailed numerical analysis of the Zugenmaier model \cite{LucasVacuum73,Lucas2013} and derived empirical scaling laws for the emission width.

A widely used approximation is
\begin{equation}
    \theta_{1/2,L} = \frac{\theta_{1/2,tr}}{\text{erf}\left(\sqrt{2/n_0^*}\right)},
    \label{EqThetaHWLucas}
\end{equation}
which correctly reproduces the transparent limit and provides a smooth interpolation across the whole molecular regime.

In the opaque limit ($n_0^* \gg 1$), this leads to the scaling
\begin{equation}
    \theta_{1/2,L} \simeq 0.63 \sqrt{n_0^*} \cdot \theta_{1/2,tr},
    \label{EqThetaHWLucasOp}
\end{equation}
highlighting the progressive degradation of collimation with increasing density.

\subsection{Discussion and limitations}

The models presented above illustrate different strategies for describing emission in the molecular regime. They can all be interpreted within the same formal framework, differing only in the prescription of the end effects and, equivalently, in the assumed density profile inside the tube.

However, none of these approaches simultaneously satisfies the following criteria:
\begin{itemize}
    \item physical transparency and clear interpretation,
    \item analytical simplicity,
    \item accuracy for both axial intensity and angular distribution,
    \item applicability across both transparent and opaque regimes.
\end{itemize}

In particular, the Clausing model is restricted to the transparent regime, the Giordmaine model fails to capture off-axis emission, and the Zugenmaier model, while accurate, remains relatively complex.

This motivates the development of a simplified analytical model that retains the essential physics of collisional transport while remaining easy to use for practical source design.

\subsection{Hanes model: secondary-emission surface}
\label{sec:second emission}

\begin{figure}[h]
	\includegraphics[width=0.99\columnwidth]{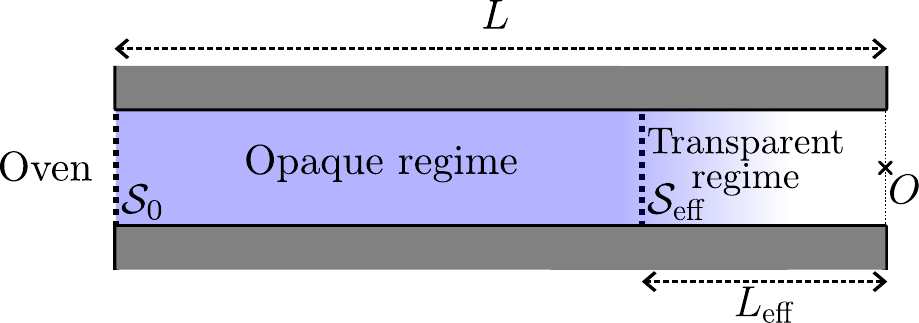}
	\caption{\label{FigHanesModel}
   Principle of the secondary-emission-surface model. In the opaque domain, the  source is replaced  by an equivalent internal emission surface
\(\mathcal{S}_{\mathrm{eff}}\) from where particles are  assumed to emerge into the remaining downstream section of the tube seen 
as a transparent emitter of effective length \(L_{\mathrm{eff}}\).
For Hanes the position is such that the remaining propagation length equals the local mean free path $ L_H=\lambda(L-L_H)$. For our HGW (Hanes-Giordmaine-Wang)  model  it is such that the axial intensity is the correct one (given by the GW model) 	so such that $L_{HGW} = L \cdot A_G(n_0^*)$.}
\end{figure}

The models discussed above rely on the direct evaluation of the  profiles once the end effects $\zeta_0$ and $\zeta_1$ have been prescribed. The model developed by Hanes \cite{HanesJAP60} follows a different philosophy. Instead of explicitly computing all source and attenuation terms, it introduces a physically intuitive secondary-emission-surface picture as illustrated by the Fig. \ref{FigHanesModel} \cite{HanesJAP60,BeijerinckJAP75}.

The starting point is the following observation. In the opaque molecular regime, the density decreases along the tube when moving toward the exit. Consequently, the local mean free path increases as
\begin{equation}
    \lambda(z) \propto \frac{1}{n(z)}.
\end{equation}
Thus, even if the flow is opaque over most of the tube, there may exist a downstream region close to the exit where the remaining path to vacuum becomes comparable to, or smaller than, the local mean free path.

Based on this idea  Hanes proposed a simple picture with atoms effectively emitted into vacuum originating from a plane located at a distance $L_H$ from the exit.
The final part of the tube, of length $L_H$, is then treated as a transparent collimating tube.

To determine $L_H$, Hanes used the simplest linear density profile, identical to the Giordmaine--Wang boundary conditions,
\begin{equation}
    \zeta_0=0, \qquad \zeta_1=1.
\end{equation}
The transition to a locally transparent region is then imposed through the condition
\begin{equation}
    \lambda(L-L_H)=L_H.
\end{equation}
With the reduced density $n_0^*=L/\lambda(z=0)$, this leads to
\begin{align}
	L_H &= L\,A_H(n_0^*), \label{EqDefLH} \\
	 n_H = n(z=L-L_H) &= n_0\,A_H(n_0^*), \label{EqDefNH}
\end{align}
with
\begin{equation}
	A_H(n_0^*) =
	\begin{cases}
		1/\sqrt{n_0^*}, & n_0^* > 1,\\
		1, & n_0^* \leq 1.
	\end{cases}
	\label{EqDefAHanes}
\end{equation}

This construction ensures continuity at $n_0^*=1$, where $L_H=L$ and $n_H=n_0$. For $n_0^*>1$, the effective emitting surface moves progressively toward the tube exit.

Using the transparent-tube result for the downstream part of the tube, Hanes obtains
\begin{align}
	I_H(0) &= \frac{n_H \bar va^2}{4}
	= I_{TW}(0)\,A_H(n_0^*),
	\label{EqIAxialHanes} \\
	f_H(\theta,n_0^*,\Gamma)
	&= f_{Clau}(\theta,\Gamma_H),
	\label{EqAngProfHanes}
\end{align}
where the effective aspect ratio is
\begin{equation}
	\Gamma_H = \frac{L_H}{2a}
	= A_H(n_0^*)\,\Gamma.
	\label{EqDefGammaHanes}
\end{equation}

The emission width is therefore immediately obtained from the transparent long-tube scaling:
\begin{eqnarray}
	\theta_{1/2,H}
	&=&
	\frac{\theta_{1/2,tr}}{A_H(n_0^*)} \\
	&\overset{\mathrm{opaque}}{=}& 
	0.84\,\frac{\sqrt{n_0^*}}{\Gamma}.
	\label{EqThetaHWHanes}
\end{eqnarray}
where the opaque overset indicate the case where $n_0^*>1$.
One of the main strength of the Hanes model is that with a very simple physical construction, it recovers the expected broadening of the beam as $\sqrt{n_0^*}$ in the opaque molecular regime.

However, the model also has important quantitative limitations. The criterion $\lambda(L_H)=L_H$ is only a rough estimate of where transparent propagation begins. It does not account neither for the fact that the mean free path continues to vary along the subsequent trajectory, nor for the gradual nature of the transition between transparent and opaque behavior \cite{FloryJAP93}. As a result, the model imposes an abrupt change at $n_0^*=1$, whereas in reality the transition is smooth and extends over an intermediate region.

\begin{figure}[h]
    \centering
    \begin{tikzpicture}
    \begin{semilogxaxis}[
        width=0.99\columnwidth,
        height=0.62\columnwidth,
        xlabel={$n_0^*$},
        ylabel={$\frac{A_Z-A_{GW(H)}}
	{A_Z}.$},
        xmin=0.01,
        xmax=100,
        ylabel style={
            at={(axis description cs:0.05,1)}, % move RIGHT (closer)
             anchor=south,
             rotate=270
        },
        grid=major,
        line width=1pt,
        tick style={black},
        label style={font=\small},
        tick label style={font=\small},
        legend style={
            font=\small,
            draw=none,
            at={(0.03,0.97)},
            anchor=north west
        },
        unbounded coords=jump,
    ]

    % --- Red curve (cut at n0* = 10)
    \addplot[
        thick,
        red,
        restrict expr to domain={\thisrowno{0}}{0.01:10}
    ]
    table[
        col sep=comma,
        x index=0,
        y index=1
    ] {Ax_Int.csv};

    % --- Green curve (DRAWN AFTER → appears on top)
    \addplot[
        thick,
        dashed,
        green!60!black
    ]
    table[
        col sep=comma,
        x index=0,
        y index=2
    ] {Ax_Int.csv};

    % --- Blue curve
    \addplot[
        thick,
        blue
    ]
    table[
        col sep=comma,
        x index=0,
        y index=3
    ] {Ax_Int.csv};

    \legend{$GW(\Gamma=10$), $GW(\Gamma=100)$, $H(\Gamma=100)$}

    \end{semilogxaxis}
    \end{tikzpicture}
    \caption{\label{FigCompAxialI}
    Relative deviations for the reduced axial intensities of the Giordmaine model (for $\Gamma=10$ and $\Gamma=100$) and the Hanes model as compared to the Zugenmaier model.}
\end{figure}
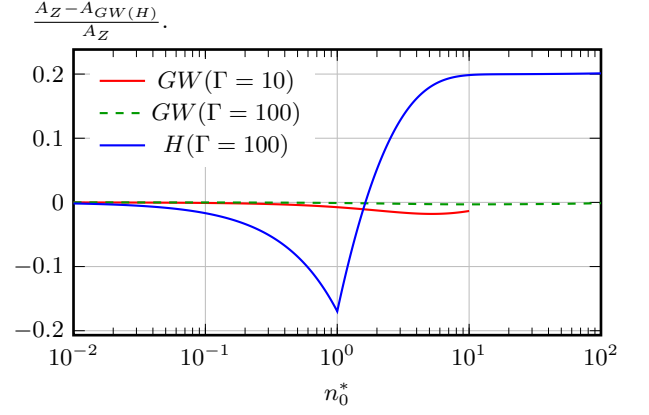

To quantify this limitation, we compare the reduced axial intensity predicted by the  Giordmaine--Wang  and the Hanes models with that of the Zugenmaier reference model. 
Figure~\ref{FigCompAxialI} shows the relative deviations \(
	\frac{A_Z(n_0^*,\Gamma)-A_{GW(H)}(n_0^*)}
	{A_Z(n_0^*,\Gamma)}
\). Since $A_Z$ depends weakly on $\Gamma$ through the end-effect coefficient $\zeta_0$, two representative aspect ratios are shown. For each case, the comparison is restricted to the molecular-flow domain, $n_0^*\leq \Gamma$.

The Giordmaine--Wang expression remains very close to the Zugenmaier prediction, with deviations typically below a few percent for $\Gamma\geq10$. This confirms that the axial intensity is only weakly affected by the precise treatment of end effects. In contrast, the Hanes prediction exhibits much larger deviations. It underestimates the axial intensity in the transition region and overestimates it in the opaque regime, with errors reaching the order of $20\%$.

The same limitation appears in the predicted angular width. Although Eq.~\eqref{EqThetaHWHanes} recovers the correct $\sqrt{n_0^*}$ scaling, its numerical prefactor is too large. Deep in the opaque regime, the Hanes width can exceed the Giordmaine--Wang or Lucas predictions by $50\%$ or more \cite{Lucas2013}. Thus, the Hanes model is physically illuminating but not sufficiently accurate for quantitative source design.

\section{Analytical  toy model}
\label{SecAnMod}

These limitations show that, although Clausing, Giordmaine--Wang, Zugenmaier, and Hanes each capture important aspects of tube emission, none provides a simple, continuous, and quantitatively reliable analytical description of both transparent and opaque molecular regimes, motivating the development of a \(10\%\)-accurate toy model.
Therefore,
in this section, we describe our proposal for the analytical model of emission. For this elaboration, we restrain the model application to well-collimated sources, and specify the associated criterion $\theta_{1/2} \lesssim 0.1$ rad. In the transparent regime Eq.~\eqref{EqThetaHWTr} readily yields the quantitative criterion $\Gamma \gtrsim 10$ for the long tube designation. To estimate the effect of opacity on $\theta_{1/2}$, we use the scaling law unveiled by Lucas' calculations via Eq.~\eqref{EqThetaHWLucasOp}, in the high opacity limit $\theta_{1/2,op} \sim \sqrt{n_0^*}/\Gamma$. A well-collimated source therefore ensures the following limitations:
\begin{equation}
	\begin{split}
		\Gamma &\gtrsim 10 \\
		n_0^* &\lesssim \left(\frac{\Gamma}{10}\right)^2 \label{EqWellCollCriteria}
	\end{split}
\end{equation}
\noindent Note that the independent restriction to the molecular regime imposes $n_0^* \leq \Gamma$, so that Eqs.~\eqref{EqWellCollCriteria} set a constraint on $n_0^*$ only for $\Gamma\leq 100$.

\subsection{Model proposal}

The discrepancies observed for the Hanes model suggest that the
secondary-emission-surface concept is physically meaningful, but that its
quantitative implementation needs to be revised. We therefore introduce a simple
analytical toy model, hereafter referred to as the \textit{HGW} model, designed to
retain the intuitive structure of Hanes' construction while correcting its main
limitations: the abrupt transition between regimes, the inaccurate prediction of
the axial intensity, and the lack of flux consistency.

The model is still based on the idea of an internal secondary emission surface,
denoted \(\mathcal{S}_{\mathrm{HGW}}\), located at a distance
\(L_{\mathrm{HGW}}\) upstream from the tube exit. This surface is assigned the
local density
\(    n_{\mathrm{HGW}} = n(z=L-L_{\mathrm{HGW}}).
\)
However, unlike in the original Hanes model, the position of
\(\mathcal{S}_{\mathrm{HGW}}\) is not determined from a local mean-free-path
criterion. Instead, it is chosen so that the model reproduces the
Giordmaine--Wang reduced axial intensity \(A_{\mathrm{GW}}\)
[Eq.~\eqref{EqDefAGiordmaine}], which provides a reliable reference for the
centerline intensity over the range of interest, as illustrated in
Fig.~\ref{FigCompAxialI}.

This constraint uniquely fixes the effective density, and therefore the position of the secondary surface within the linear-density picture.

\noindent The axial intensity predicted by the HGW model is $I_{HGW}\left(0\right)=n_{HGW} \bar v a^2/4$. Imposing $I_{HGW}\left(0\right)/I_{TW}\left(0\right)=A_G(n_0^*)$ and assuming the linear density evolution (Eq. \ref{EqScDensityVsZOp}) results in:

\begin{align}
	L_{HGW} &= L  \, A_G(n_0^*) \label{EqDefLHG} \\
	n_{HGW} &= n_0  \, A_G(n_0^*) \label{EqDefNHG}
\end{align}

Our model did not assume an abrupt change in physical behavior, since the transition between regimes remain smooth; we only state that the final beam 
controlled by $A_G(n_0^*)$,
can be understood is if it was emitted, at density  $n_{HGW}$ in a transparent way by the surface located at  distance $L_{HGW}$ from the exit.

Once the effective source is defined, the angular distribution follows directly from transparent-regime physics.
The HGW angular profiles are deduced from the axial intensity $A_G$ and the Clausing profiles for a collimating tube of effective length $L_{HGW}$, thus presenting a simple analytical formulation. Below we summarize these quantities in the same manner as for the Hanes model:
\begin{align}
	I_{HGW}\left(0\right) &= \frac{n_{HGW} \bar v a^2}{4} = I_{TW}\left(0\right) \cdot A_G(n_0^*)\label{EqIAxialHG}\\
	f_{HGW}\left(\theta,n_0^*,\Gamma\right) &= f_{Clau}\left(\theta,\Gamma_{HGW}\right) \label{EqAngProfHG}\\
 	\text{with: } \Gamma_{HGW} &= \frac{L_{HGW}}{2a} = \mathcal{A}_G\left(n_0^*\right) \cdot \Gamma \label{EqDefGammaHG}
\end{align}

Opacity therefore couples intensity reduction and angular broadening through a single parameter $A_G(n_0^*)$.

\subsection{Model predictions}
\label{sec:prediction}

Here, we discuss the model predictions and compare them with the reference Zugenmaier model. For clarity, we distinguish two emission-angle ranges based on the emission width $\theta_{1/2}$: near-axis emission for $\theta \lesssim \theta_{1/2}$, including axial emission; and off-axis emission defined by $\theta_{1/2} \lesssim \theta \leq \pi/2$.

By nature, the HGW model predicts the reference reduced axial intensity $A_G(n_0^*)$ \cite{GiordmaineJAP60}. We thus consider that this quantity is perfectly captured by the HGW model and does not require further inspection.

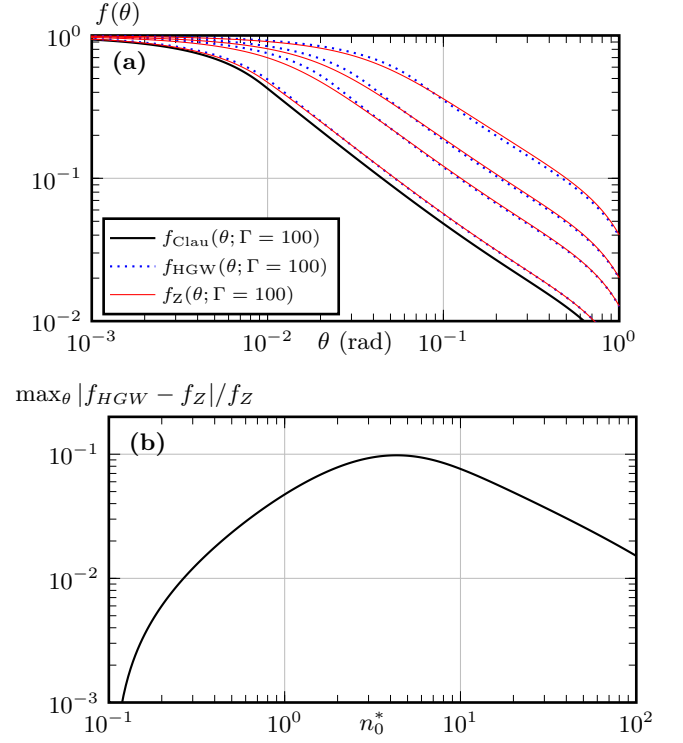
\begin{figure}[t]
    \centering

    % =========================
    % (a) Main plot
    % =========================
    \begin{tikzpicture}
    \begin{loglogaxis}[
        width=0.99\columnwidth,
        height=0.62\columnwidth,
        xlabel={$\theta\ \mathrm{(rad)}$},
        ylabel={$f(\theta)$},
        xmin=0.001,
        xmax=1,
        ymin=0.01,
        ymax=1,
        xlabel style={
            at={(axis description cs:0.5,0)}, % move UP (closer to axis)
            anchor=north
        },
        ylabel style={
            at={(axis description cs:0.05,1)}, % move RIGHT (closer)
             anchor=south,
             rotate=270
        },
        grid=major,
        line width=1pt,
        tick style={black},
        label style={font=\small},
        tick label style={font=\small},
        legend style={
            font=\scriptsize,
            draw=black,
            fill=white,
            at={(0.02,0.02)},
            anchor=south west,
            inner sep=2pt
        },
        legend cell align=left
    ]

    % Clausing profile
    \addplot[
        thick,
        black
    ]
    table[
        col sep=comma,
        x index=0,
        y index=9
    ] {fHGZug.csv};
    \addlegendentry{$f_{\mathrm{Clau}}(\theta;\Gamma=100)$}

    % HGW profiles
    \addplot[
        thick,
        dotted,
        blue
    ]
    table[
        col sep=comma,
        x index=0,
        y index=1
    ] {fHGZug.csv};
    \addlegendentry{$f_{\mathrm{HGW}}(\theta;\Gamma=100)$}

    \addplot[thick, dotted, blue, forget plot]
    table[col sep=comma, x index=0, y index=3] {fHGZug.csv};

    \addplot[thick, dotted, blue, forget plot]
    table[col sep=comma, x index=0, y index=5] {fHGZug.csv};

    \addplot[thick, dotted, blue, forget plot]
    table[col sep=comma, x index=0, y index=7] {fHGZug.csv};

    % Zugenmaier profiles
    \addplot[
        thin,
        red
    ]
    table[
        col sep=comma,
        x index=0,
        y index=2
    ] {fHGZug.csv};
    \addlegendentry{$f_{\mathrm{Z}}(\theta;\Gamma=100)$}

    \addplot[thin, red, forget plot]
    table[col sep=comma, x index=0, y index=4] {fHGZug.csv};

    \addplot[thin, red, forget plot]
    table[col sep=comma, x index=0, y index=6] {fHGZug.csv};

    \addplot[thin, red, forget plot]
    table[col sep=comma, x index=0, y index=8] {fHGZug.csv};

    % panel label
    \node[font=\small, anchor=north west] at (rel axis cs:0.02,0.98) {\textbf{(a)}};

    \end{loglogaxis}
    \end{tikzpicture}

    \vspace{0.4em}

    % =========================
    % (b) Former inset as separate plot
    % =========================
    \begin{tikzpicture}
    \begin{loglogaxis}[
        width=0.99\columnwidth,
        height=0.62\columnwidth,
        xlabel={$n_0^*$},
        ylabel={$\max_\theta |f_{HGW}-f_Z|/f_Z$},
        xlabel style={
            at={(axis description cs:0.5,0)}, % move UP (closer to axis)
            anchor=north
        },
        ylabel style={
            at={(axis description cs:0.05,1)}, % move RIGHT (closer)
             anchor=south,
             rotate=270
        },
        xmin=0.1,
        xmax=100,
        ymin=0.001,
        ymax=0.2,
        grid=major,
        line width=1pt,
        tick style={black},
        label style={font=\small},
        tick label style={font=\small},
    ]

    \addplot[
        thick,
        black
    ]
    table[
        col sep=comma,
        ignore chars={"},
        skip first n=1,
        x index=0,
        y index=1
    ] {RelDevHGZug.csv};

    % panel label
    \node[font=\small, anchor=north west] at (rel axis cs:0.02,0.98) {\textbf{(b)}};

    \end{loglogaxis}
    \end{tikzpicture}

    \caption{\label{FigCompProfZugHG}
    \textbf{(a)} Comparison between the angular profiles predicted by the Zugenmaier model (thin red lines) and the HGW model (thick dotted blue lines), for $\Gamma=100$ and selected values of the reduced oven density $n_0^*=1,10,25,100$. The Clausing profile of the transparent regime is shown in thick black.
    \textbf{(b)} Maximum absolute value of the relative deviation between the HGW and Zugenmaier angular profiles as a function of $n_0^*$.}
\end{figure}

Following the definitions of $f_{HGW}$ and $\Gamma_{HGW}$, it is obvious that the HGW model predicts the Clausing angular profiles in the transparent regime. We thus focus on the predictions of the HGW model in the opaque regime, and compare it to those of the reference Zugenmaier model. Figure \ref{FigCompProfZugHG} (a) displays several HGW and Zugenmaier profiles computed for $\Gamma=100$ and a series of values of $n_0^*$ in the opaque regime ($1 \leq n_0^* \leq 100$).

The angular range $\theta \leq 1$ rad and the logarithmic scale are chosen to highlight differences between the $f_Z$ and $f_{HGW}$ predictions. Both models yield nearly indistinguishable results at low and large emission angles, with only slight discrepancies in the intermediate range. These differences are quantified in the Fig. \ref{FigCompProfZugHG} (b) by the maximum absolute relative deviation, defined as the peak max  of $\left|f_{HGW}(\theta)-f_Z(\theta)\right|/f_Z(\theta)$, over $0 \leq \theta < \pi/2$, as a function of $n_0^*$. The deviations are greatest for $1 \lesssim n_0^* \lesssim 10$, that is, in the region just above the transition between the transparent and opaque regimes. Qualitatively, the cusp-shaped functional form of the Clausing profile (black line) is preserved in the opaque regime \cite{GiordmaineJAP60,HanesJAP60}, consistent with predictions for a well-collimated source.

In the same manner as for the Hanes model, the HGW model predicts the emission width with the following simple formula:
\begin{equation}
	\theta_{1/2,HGW} = \frac{0.84}{\Gamma_{HGW}} = \frac{\theta_{1/2,tr}}{A_{\rm GW}(n_0^*)} \label{EqThetaHWHG}
\end{equation}
\noindent Contrary to Eq.~\eqref{EqThetaHWHanes} which application involves a conditional definition, Eq.~\eqref{EqThetaHWHG} is fully analytical over all the density range of the molecular flow and provides a smooth transition between the flow regimes.

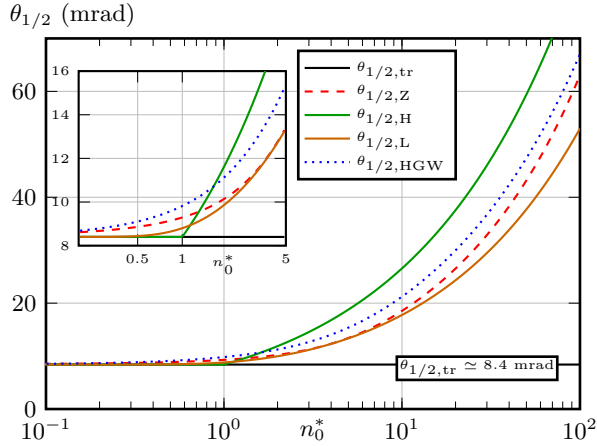
\begin{figure}[h]
    \centering
    \begin{tikzpicture}

    % =========================
    % Main plot
    % =========================
    \begin{semilogxaxis}[
        name=mainplot,
        width=1\columnwidth,
        height=0.75\columnwidth,
        xlabel={$n_0^*$},
        ylabel={$\theta_{1/2}\ \mathrm{(mrad)}$},
        xlabel style={
            at={(axis description cs:0.5,0.)}, % move UP (closer to axis)
            anchor=north
        },
        ylabel style={
            at={(axis description cs:0.05,1)}, % move RIGHT (closer)
             anchor=south,
             rotate=270
        },
        xmin=0.1,
        xmax=100,
        ymin=0,
        ymax=70,
        grid=major,
        line width=1pt,
        tick style={black},
        label style={font=\small},
        tick label style={font=\small},
        legend style={
            font=\tiny,
            draw=black,
            fill=white,
            at={(0.47,0.97)},
            anchor=north west,
            inner sep=2pt
        },
        legend cell align=left
    ]

    % Transparent regime
    \addplot[
        thick,
        black
    ]
    table[
        col sep=comma,
        ignore chars={"},
        skip first n=1,
        x index=0,
        y expr=1000*\thisrowno{1}
    ] {HWHM.csv};
    \addlegendentry{$\theta_{1/2,\mathrm{tr}}$}

    % Zugenmaier
    \addplot[
        thick,
        dashed,
        red
    ]
    table[
        col sep=comma,
        ignore chars={"},
        skip first n=1,
        x index=0,
        y expr=1000*\thisrowno{2}
    ] {HWHM.csv};
    \addlegendentry{$\theta_{1/2,\mathrm{Z}}$}

    % Hanes
    \addplot[
        thick,
        green!60!black
    ]
    table[
        col sep=comma,
        ignore chars={"},
        skip first n=1,
        x index=0,
        y expr=1000*\thisrowno{3}
    ] {HWHM.csv};
    \addlegendentry{$\theta_{1/2,\mathrm{H}}$}

    % Lucas
    \addplot[
        thick,
        orange!80!black
    ]
    table[
        col sep=comma,
        ignore chars={"},
        skip first n=1,
        x index=0,
        y expr=1000*\thisrowno{4}
    ] {HWHM.csv};
    \addlegendentry{$\theta_{1/2,\mathrm{L}}$}

    % HGW
    \addplot[
        thick,
        dotted,
        blue
    ]
    table[
        col sep=comma,
        ignore chars={"},
        skip first n=1,
        x index=0,
        y expr=1000*\thisrowno{5}
    ] {HWHM.csv};
    \addlegendentry{$\theta_{1/2,\mathrm{HGW}}$}

    % Giordmaine
    %\addplot[
     %   thick,
      %  solid,
       % purple
%    ]
 %   table[
  %      col sep=comma,
   %     ignore chars={"},
    %    skip first n=1,
     %   x index=0,
      %  y expr=1000*\thisrowno{6}
    %] {HWHM.csv};
    %\addlegendentry{$\theta_{1/2,\mathrm{G}}$}
    
    \node[
        font=\tiny,
        draw=black,
        fill=white,
        inner sep=1pt,
        anchor=south east
    ] at (rel axis cs:0.96,0.075) {$\theta_{1/2,\mathrm{tr}}\simeq 8.4\ \mathrm{mrad}$};

    \end{semilogxaxis}

    % =========================
    % Inset
    % =========================
    \begin{semilogxaxis}[
        at={(mainplot.north west)},
        anchor=north west,
        xshift=0.05\columnwidth,
        yshift=-0.05\columnwidth,
        width=0.5\columnwidth,
        height=0.45\columnwidth,
        xmin=0.2,
        xmax=5,
        ymin=8,
        ymax=16,
        grid=major,
        line width=0.8pt,
        tick style={black},
        tick label style={font=\tiny},
        label style={font=\tiny},
        xtick={0.5,1,5},
        xticklabels={0.5,1,5},
        xlabel={$n_0^*$},
        ylabel={},
        xlabel style={
            font=\tiny,
            at={(axis description cs:0.7,0)},
            anchor=north
        },
        clip=true
    ]

    % Transparent regime
    \addplot[
        thick,
        black
    ]
    table[
        col sep=comma,
        ignore chars={"},
        skip first n=1,
        x index=0,
        y expr=1000*\thisrowno{1}
    ] {HWHM_inset.csv};

    % Zugenmaier
    \addplot[
        thick,
        dashed,
        red
    ]
    table[
        col sep=comma,
        ignore chars={"},
        skip first n=1,
        x index=0,
        y expr=1000*\thisrowno{2}
    ] {HWHM_inset.csv};

    % Hanes
    \addplot[
        thick,
        green!60!black
    ]
    table[
        col sep=comma,
        ignore chars={"},
        skip first n=1,
        x index=0,
        y expr=1000*\thisrowno{3}
    ] {HWHM_inset.csv};

    % Lucas
    \addplot[
        thick,
        orange!80!black
    ]
    table[
        col sep=comma,
        ignore chars={"},
        skip first n=1,
        x index=0,
        y expr=1000*\thisrowno{4}
    ] {HWHM_inset.csv};

    % HGW
    \addplot[
        thick,
        dotted,
        blue
    ]
    table[
        col sep=comma,
        ignore chars={"},
        skip first n=1,
        x index=0,
        y expr=1000*\thisrowno{5}
    ] {HWHM_inset.csv};

        % Giordmaine
    %\addplot[
     %   thick,
      %  solid,
       % purple
    %]
    %table[
     %   col sep=comma,
      %  ignore chars={"},
       % skip first n=1,
       % x index=0,
       % y expr=1000*\thisrowno{6}
    %] {HWHM_inset.csv};

    \end{semilogxaxis}

    \end{tikzpicture}
    \caption{\label{FigAngWidthOp}
    Half width at half maximum $\theta_{1/2}$ predicted by the different emission models for $\Gamma=100$, as a function of the reduced oven density $n_0^*$. The inset shows a magnified view of the transition region region.}
\end{figure}

Figure~\ref{FigAngWidthOp} plots the value of $\theta_{1/2}$ as a function of $n_0^*$ for $\Gamma=100$, as predicted by all the emission models which capture the opaque regime. The inset is a zoom in the transition region. We provide a detailed observation of the various predictions as compared to the reference Zugenmaier emission width $\theta_{1/2,Z}$, displayed with the thick red dashed line. Thus  numerical computation of $\theta_{1/2,Z}$ require much less computational resources than the  direct analytical equations describing other emission widths.

The emission width $\theta_{1/2,L}$ obtained by Lucas (Eq.\eqref{EqThetaHWLucas}, orange line) is a phenomenological guess for an analytical description of $\theta_{1/2,Z}$: it captures both regimes of flow and connects them smoothly, but among the analytical predictions, appears to yield the largest underestimate of $\theta_{1/2}$ deep in the opaque regime, althoug still reasonable (around 15\%). Concerning the simplest formula provided by the Hanes model (Eq.~\eqref{EqThetaHWHanes}, green line), the abrupt transition between regimes is clearly visible at $n_0^*=1$ (inset); deep in the opaque regime, it yields the largest overestimate exceeding 30\%. Finally, the HGW model (Eq.~\eqref{EqThetaHWHG}, thick blue dotted line) appears to yield the largest overestimate of $\theta_{1/2}$ at the transition between regimes (inset), but the most accurate analytical prediction with rising opacity: the relative deviation is naturally negligible in the transparent regime, close to 6\% for $n_0^*=1$, peaks at 15\% for $n_0^* \simeq 7$ and then decreases back to 6\% for $n_0^*=100$.
Thus, unlike the Hanes model, which predicts an overestimated width due to its abrupt transition, the present expression provides a smooth evolution consistent with the Giordmaine--Wang scaling and the broadening directly reflects the reduction of the effective emitting length.

The physical consistency of the model with respect to total flux conservation constitutes a necessary criterion for assessing its reliability \cite{LucasVacuum73,OlanderJAP70III,Lucas2013}. The intensity prescribed by a given model should, upon integration over the solid angle, yield the Knudsen flow predicted by Eq.~\eqref{EqTotFluxGeneral}. It should be noted that the exact form of the tube impedance depends on the model considered: for all models based on the transparent-flow description (Clausing, Hanes, HGW), $W_{Clau}$ is taken as an accurate estimate, whereas the Zugenmaier model relies on the more elaborate expression $W_Z$.

We define $C$ as the integrated flux emitted by the oven, normalized by the impedance. It is therefore clear that an idealized perfectly accurate model implies $C=1$.

\begin{equation}
	\mathcal{C} = \frac{2}{W} \int_0^{\pi/2} A\left(n_0^*\right)f(\theta)\sin \theta d\theta \label{EqTestIntTotFlux}
\end{equation}

\begin{figure}[h]
    \centering
    \begin{tikzpicture}
    \begin{semilogxaxis}[
        width=0.85\columnwidth,
        height=0.62\columnwidth,
        scale only axis,
        xlabel={$n_0^*$},
        ylabel={$C$},
        xmin=0.1,
        xmax=100,
        grid=major,
        line width=1pt,
        tick style={black},
        label style={font=\small},
        tick label style={font=\small},
        xlabel style={
            at={(axis description cs:0.5,0.)},
            anchor=north
        },
        ylabel style={
            at={(axis description cs:-0.05,1.)},
            anchor=south,
            rotate=270
        },
        legend style={
            font=\scriptsize,
            draw=black,
            fill=white,
            at={(0.02,0.5)},
            anchor=north west,
            inner sep=2pt
        },
        legend cell align=left
    ]

    % Zugenmaier (Gamma = 100)
    \addplot[
        thick,
        red
    ]
    table[
        col sep=comma,
        ignore chars={"},
        skip first n=1,
        x index=0,
        y index=1
    ] {TotalFlux_100_LT.csv};
    \addlegendentry{Zugenmaier $(\Gamma=100)$}

    % Hanes (Gamma = 100)
    \addplot[
        thick,
        dashed,
        black
    ]
    table[
        col sep=comma,
        ignore chars={"},
        skip first n=1,
        x index=0,
        y index=2
    ] {TotalFlux_100_LT.csv};
    \addlegendentry{Hanes $(\Gamma=100)$}

    % HGW (Gamma = 100)
    \addplot[
        thick,
        dotted,
        blue
    ]
    table[
        col sep=comma,
        ignore chars={"},
        skip first n=1,
        x index=0,
        y index=3
    ] {TotalFlux_100_LT.csv};
    \addlegendentry{HGW $(\Gamma=100)$}

    % HGW (Gamma = 30) -- truncated
    \addplot[
        thick,
        dashed,
        green!60!black,
        restrict expr to domain={\thisrowno{0}}{0.1:9}
    ]
    table[
        col sep=comma,
        ignore chars={"},
        skip first n=1,
        x index=0,
        y index=4
    ] {TotalFlux_100_LT.csv};
    \addlegendentry{HGW $(\Gamma=30)$}

    % HGW (Gamma = 10) -- truncated
    \addplot[
        thick,
        dotted,
        orange!80!black,
        restrict expr to domain={\thisrowno{0}}{0.1:1}
    ]
    table[
        col sep=comma,
        ignore chars={"},
        skip first n=1,
        x index=0,
        y index=5
    ] {TotalFlux_100_LT.csv};
    \addlegendentry{HGW $(\Gamma=10)$}

    \end{semilogxaxis}
    \end{tikzpicture}
    \caption{\label{FigConsTestTotFluxOp}
    Value of $C$ as a function of the reduced oven density $n_0^*$, for the Zugenmaier, Hanes and HGW models for different aspect ratios. The curves are presented in accordance with restriction to well-collimated sources (Eq. \ref{EqWellCollCriteria}).}
\end{figure}
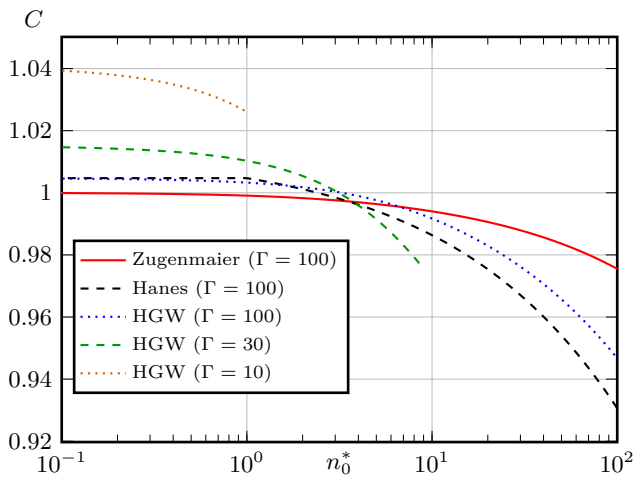

Figure~\ref{FigConsTestTotFluxOp} shows \(C\) as a function of \(n_0^*\) for aspect ratios ranging from \(\Gamma=10\) to \(\Gamma=100\), representative of the lower and upper bounds of the long-tube regime. The HGW model is shown over this range, while the Hanes and Zugenmaier models are included for \(\Gamma=100\) as reference comparisons.

For $\Gamma=100$, the HGW model appears consistent throughout the molecular flow regime, with a maximum deviation of 5\% at maximum opacity, $n_0^*=100$, aided by the choice $\alpha = W_{LT}/2$. The Hanes model exhibits a slightly larger maximum deviation of 7\%, which remains surprisingly low compared with the much larger deviations observed for other quantities such as the axial intensity or the emission width. This highlights the following strength of the secondary emission surface approach: its inherent approximate conservation of the integrated flux \cite{HanesJAP60}.

As aspect ratio decreases, one can observe an increasing overestimate of the flux by HGW in the transparent regime (up to $\sim 4\%$ for $\Gamma=10$). This can be explained by the fact that, for shorter tubes, the estimate of the end effects using $\alpha = W_{LT}/2$ appears less accurate. The same also applies to the choice $W = W_{Clau}$ for the tube impedance \cite{ClausingAdP32}.

\subsection{Discussion and perspectives}

From the presented flow characteristics, we can draw the following conclusions on the emission toy model predictions:

The HGW reproduces the near-axis emission of the reference Zugenmaier description with high fidelity, by construction for the axial intensity and, more generally, for the low-angle part of the angular profile. This can be understood by noting that near-axis emission is dominated by direct transmission from the oven and by emission from large portions of the inner tube wall, for which end effects remain small and have a negligible overall influence. The remaining discrepancies are mainly observed in the transition region between the transparent and opaque regimes. In this zone, the secondary-emission-surface approximation becomes less accurate, as the source opacity is not yet large enough for the far-field emission to originate from a well-defined portion of the tube wall effectively delimited by such a surface.

The off-axis emission is primarily governed by the density distribution in restricted regions of the tube close to its exit. The HGW preserves the correct qualitative behavior at large angles and remains close to the Zugenmaier profiles. This is a direct consequence of the model construction, which is based on the Clausing profiles and therefore remains valid for arbitrary emission angles. These results indicate that the secondary-emission-surface picture captures the dominant geometrical contribution to off-axis emission for long tubes operating in the molecular regime, while avoiding the computational cost associated with the evaluation of opaque Olander--Kruger type expressions. Within the accuracy target adopted in this work, the HGW thus provides a reliable analytical proxy for source design studies, particularly at higher opacity.

Discussing the perspective of application of the HGW model, we want to highlight its potential for the description of the flow properties within the collisional flow regime. In general, the secondary-emission-surface approach remains valid whenever the transparent regime is regained in a small portion of the tube connected to its output. Naturally, with the evolution of opacity, one can expect transition from collisional ($n_0^*>\Gamma$) to molecular ($n_0^*<\Gamma$) regime, and re-establishment of the conditions viable for the application of the HGW model. However, it should be noted that proper modeling of the collisional part of the flow close to the input of the tube will be necessary to obtain a correct value of $n_0^*$ for the remaining part of the tube, where the molecular regime revives.

\begin{table*}[t]
\centering
\scriptsize
\setlength{\tabcolsep}{5pt}
\renewcommand{\arraystretch}{1.45}

\begin{tabular}{|l||l|l|}
\hline
& thin-wall aperture
& HGW tube model \\
\hline
\hline
Geometry
&
\makecell[l]{
aperture diameter \(d\)
}
&
\makecell[l]{
tube diameter \(d\), length \(L\)\\
\(\Gamma=L/d\), \qquad \(n_0^*=L/\lambda\)
}
\\
\hline

Effective emission source
&
\makecell[l]{
physical aperture\\
density \(n_0\)
}
&
\makecell[l]{
\(A_{\rm GW}(n_0^*)=
\dfrac{\sqrt{\pi}}{2}
\sqrt{\dfrac{2}{n_0^*}}\,
\operatorname{erf}\!\left(\sqrt{\dfrac{n_0^*}{2}}\right)\)\\
\(L_{\rm HGW}=L \, A_{\rm GW}(n_0^*)\)\\
\(\Gamma_{\rm HGW}=\Gamma \,  A_{\rm GW}(n_0^*)\)
}
\\
\hline

Axial intensity
&
\makecell[l]{
\(I_{\rm TW}(0)=\dfrac{n_0\bar v d^2}{16}\)
}
&
\makecell[l]{
\(I_{\rm HGW}(0)=I_{\rm TW}(0)A_{\rm GW}(n_0^*)\)
}
\\
\hline

Angular intensity
&
\makecell[l]{
\(I_{\rm TW}(\theta)=I_{\rm TW}(0)\cos\theta\)
}
&
\makecell[l]{
\(I_{\rm HGW}(\theta)=I_{\rm HGW}(0) f_{HGW}(\theta)\) \, \\
\( f_{\rm HGW}(\theta)=
\begin{cases}
\displaystyle
\cos\theta\left[
\alpha+\frac{2}{\pi}
\left\{
(1-\alpha)R(q)
+\frac{2(1-2\alpha)}{3q}
\left[1-(1-q^2)^{3/2}\right]
\right\}
\right],
& q\leq1,\\[2ex]
\displaystyle
\cos\theta\left[
\alpha+\frac{4(1-2\alpha)}{3\pi q}
\right],
& q\geq1.
\end{cases}\)\\
\ \ \ with \(q=\Gamma_{\rm HGW} \tan\theta\), \,
\(\alpha=2/(3\Gamma_{\rm HGW})\), \,
\(R(q)=\arccos q-q\sqrt{1-q^2}\)
}
\\
\hline

Total flux
&
\makecell[l]{
\(\dot N_{\rm TW}=\pi I_{\rm TW}(0)\)\\
\(\dot N_{\rm TW}=\dfrac{\pi n_0\bar v d^2}{16}\)
}
&
\makecell[l]{
physical tube flux:\\
\(\dot N_{\rm tube}\simeq \dot N_{\rm TW}W_{\rm Clau}(\Gamma)\)\\
\(W_{\rm Clau}(\Gamma)=\dfrac{4}{3\Gamma+4}\)\\[0.3em]
angular ansatz gives:\\
\(\dot N_{\rm ang}=\dot N_{\rm TW}A_{\rm GW}
W_{\rm Clau}(\Gamma_{\rm HGW})\)
}
\\
\hline

Angular half-width
&
\makecell[l]{
\(\theta_{1/2,\rm TW}=\pi/3\)
}
&
\(\theta_{1/2,\rm HGW}\simeq
\dfrac{0.84}{\Gamma_{\rm HGW}}\)
\\

\hline
\end{tabular}
\caption{
Self-contained summary of the thin-wall aperture (used as reference) and of the HGW  model for an oven and a tube at temperature \(T\).
Here \(n_0\) is the reservoir density, \(d\) the tube diameter, \(L\) the tube length,
\(\Gamma=L/d\), \(\bar v=\sqrt{8k_B T/(\pi m)}\), and
\(\lambda=(\sqrt{2}\sigma n_0)^{-1}\), with \(\sigma=\pi d_{\rm kin}^2\).
The reduced oven density is \(n_0^*=L/\lambda\).
The function \(A_{\rm GW}(n_0^*)\) is the Giordmaine--Wang reduced axial intensity,
\(A_{\rm GW}=I(0)/I_{\rm TW}(0)\). It is used  to set the position and density
of the HGW effective emission surface. 
The function
\(f_{\rm Clau}(\theta,\Gamma_{\rm HGW})\) is the normalized transparent Clausing
angular profile of a bright-wall tube  evaluated with the effective aspect ratio. It can be written as 
\(f_{\rm HGW}(\theta) =
f_{\rm Clau}(\theta,\Gamma_{\rm HGW})
\).
The total flux can be estimated either from the Knudsen/Clausing
throughput of the full tube,
\(\dot N_{\rm tube}\simeq \dot N_{\rm TW}W_{\rm Clau}(\Gamma)\), or by integrating
the HGW angular distribution,
\(\dot N_{\rm ang}=\dot N_{\rm TW}A_{\rm GW}W_{\rm Clau}(\Gamma_{\rm HGW})\).
Although \(W_{\rm Clau}\) is strictly a transparent-regime transmission
probability, this throughput remains a good approximation throughout the
opaque molecular range \(d<\lambda<L\) \cite{GiordmaineJAP60,Pauly}, where collisions mainly modify the
axial intensity and angular profile; the two estimates agree within a few percent
in the range considered here, as shown in Fig.~\ref{FigConsTestTotFluxOp}.
}
\label{TabHGWModel}
\end{table*}

\section{Conclusions}

In this work, we have theoretically investigated effusive primary-source
emission through long bright-wall collimating tubes and introduced a simple
analytical description of the flow. The proposed Hanes--Giordmaine--Wang
(HGW) model fully summarized in Table \ref{TabHGWModel} provides a compact compromise between physical transparency,
analytical simplicity, and quantitative accuracy. Its central idea is to retain
the intuitive secondary-emission-surface picture introduced by Hanes, while
choosing the effective emission surface so as to recover the accurate Giordmaine--Wang
axial intensity. In this way, the model preserves the geometrical clarity of the
transparent Clausing regime while incorporating, through a single effective
length, the dominant effect of interparticle collisions in the opaque molecular
regime.

The HGW model predicts the axial intensity by construction and gives simple
analytical expressions for the angular profile, the emission width, and the
integrated flux consistency. Compared with the more complete Zugenmaier--
Olander--Kruger description, the largest deviations are found in the transition region
between the transparent and opaque regimes, where the notion of a sharply
defined effective emission surface is necessarily approximate. These deviations
remain of the order required for preliminary source design, typically at the \(10\%\) level.

The model also provides useful design guidance. For a single tube, increasing
the aspect ratio \(\Gamma=L/d\) improves collimation. At fixed source density, this
increases the beam brightness (total flux divided by the emitting area and by the solid angle divergence \cite{LEJEUNEC1980,rhee1992refined,Brau02,HahnRSI22}),
\(
    B
    =
    \frac{\dot N}
    {(\pi d^2/4)\,\pi\theta_{1/2}^2},
\)
because the reduction of angular divergence  can
more than compensate the decrease of throughput  (roughly both $\dot N$ and $\theta_{1/2}$ scales has $1/\Gamma$). However, a single tube also
has a fundamental limitation: increasing the tube diameter is the direct way to
increase the total flux, but it simultaneously pushes the system more rapidly
toward the collisional  regimes when the source density is
raised. Thus, a single large tube is generally not the optimal architecture for a
high-flux, well-collimated molecular beam.

A more favorable strategy is to use a multichannel array \cite{HanesJAP60,BeijerinckJPE74,seccombe2001design,guevremont2000design,majumder2009generation,buckman1993spatial,hopkins2016versatile,kemp2016production,Hughes2026}, treated to first
approximation as a set of independent tubes. This allows one to combine a large
total open area with a small individual tube diameter. The latter helps maintain
the molecular-flow condition, \(d\ll \lambda\), at higher operating densities, while
the large number of channels preserves a high total flux. 
Although the most relevant figure of merit ultimately depends on the target experiment, the beam brightness is often a natural optimization criterion; another commonly used criterion, introduced by Hanes \cite{HanesJAP60}, is \((\phi/d)^{1/2}\), where \(\phi\) is the open-area fraction of the array.
Whatever choice made it supports the use of dense microcapillary, and
possibly nanocapillary, arrays for high-brightness effusive sources, provided
that the bright-wall assumption remains valid and that clogging, surface
reactivity, and fabrication constraints are controlled
\cite{karniadakis2005microflows}. 

The aspect ratio must be chosen as a compromise rather than made arbitrarily
large. A value \(\Gamma \gtrsim 10\) is enough to reach the long-tube regime,
while \(\Gamma \sim 30{-}100\) is a useful practical range when opaque molecular
operation is expected. Larger \(\Gamma\) values further reduce the divergence,
but only at the cost of a lower transmitted flux; beyond some point, the gain in collimation no
longer compensates the loss in flux.

We believe that the analytical framework proposed here will be useful as a
first-step design tool for effusive atomic and molecular sources. By reducing
the complex collisional emission problem to a small set of transparent
parameters, the HGW model can guide the choice of tube diameter, aspect ratio,
open fraction, and operating density before more detailed numerical simulations
or experimental optimization are undertaken.

\section{Acknowledgements}

This work was supported by French National Research agency (ANR), under grant ANR-22-CE47-0005 (QIPRYA project).

 A CC-BY public copyright license has been applied by the authors to the present document and will be applied to all subsequent versions up to any Author Accepted Manuscript (AAM) version arising from this submission, in accordance with the grant's open access conditions.

\bibliography{HG-Model}{}

@misc{Hughes2026,
      title={A high-flux atomic strontium oven with light-driven flux modulation}, 
      author={Kenneth M. Hughes and Jesse S. Schelfhout and Charu Mishra and Timothy Leese and Elliot Bentine and Christopher J. Foot},
      year={2026},
      eprint={2603.25567},
      archivePrefix={arXiv},
      primaryClass={physics.atom-ph},
      url={https://arxiv.org/abs/2603.25567}, 
}

@article{AdamsonVacuum88II,
	title = {The angular distribution of thermal molecular beams formed by single capillaries in the molecular flow regime},
	journal = {Vacuum},
	volume = {38},
	number = {6},
	pages = {463-467},
	year = {1988},
	issn = {0042-207X},
	doi = {https://doi.org/10.1016/0042-207X(88)90589-1},
	url = {https://www.sciencedirect.com/science/article/pii/0042207X88905891},
	author = {S Adamson and C O'Carroll and JF McGilp}
}

@article{AquilantiJCP99,
	author = {Aquilanti, V. and Ascenzi, D. and de Castro V\'itores, M. and Pirani, F. and Cappelletti, D.},
	title = {A quantum mechanical view of molecular alignment and cooling in seeded supersonic expansions},
	journal = {The Journal of Chemical Physics},
	volume = {111},
	number = {6},
	pages = {2620-2632},
	year = {1999},
	month = {08},
	issn = {0021-9606},
	doi = {10.1063/1.479537},
	url = {https://doi.org/10.1063/1.479537},
}

@article{BeijerinckJPE74,
	doi = {10.1088/0022-3735/7/1/009},
	url = {https://dx.doi.org/10.1088/0022-3735/7/1/009},
	year = {1974},
	month = {jan},
	publisher = {},
	volume = {7},
	number = {1},
	pages = {31},
	author = {H C W Beijerinck and  R G J M Moonen and  N F Verster},
	title = {Calibration of a time-of-flight machine for molecular beam studies},
	journal = {Journal of Physics E: Scientific Instruments},
}

@article{IvanovTroitskii1963,
  author  = {Ivanov, B. S. and Troitskii, V. S.},
  title   = {Formation of Directivity Patterns of Molecular Beams},
  journal = {Soviet Physics--Technical Physics},
  volume  = {8},
  pages   = {365--368},
  year    = {1963},
  note    = {Original Russian: Zhurnal Tekhnicheskoi Fiziki 33, 494--499 (1963)}
}

@article{Becker1961,
  author  = {Becker, G.},
  title   = {Zur Theorie der Molekularstrahlerzeugung mit langen Kan{\"a}len},
  journal = {Zeitschrift f{\"u}r Physik},
  volume  = {162},
  pages   = {290--312},
  year    = {1961},
  doi     = {10.1007/BF01341977}
}

@article{BeijerinckJAP75,
	author = {Beijerinck, H. C. W. and Verster, N. F.},
	title = "{Velocity distribution and angular distribution of molecular beams from multichannel arrays}",
	journal = {Journal of Applied Physics},
	volume = {46},
	number = {5},
	pages = {2083-2091},
	year = {1975},
	month = {05},
	issn = {0021-8979},
	doi = {10.1063/1.321845},
	url = {https://doi.org/10.1063/1.321845},
}

@article{BeijerinckPhysica76,
	title = {Monte-Carlo of molecular flow through a cylindrical channel},
	journal = {Physica B+C},
	volume = {83},
	number = {2},
	pages = {209-219},
	year = {1976},
	issn = {0378-4363},
	doi = {https://doi.org/10.1016/0378-4363(76)90223-0},
	url = {https://www.sciencedirect.com/science/article/pii/0378436376902230},
	author = {H.C.W. Beijerinck and M.P.J.M. Stevens and N.F. Verster},
}

@article{BernasekPSS75,
	title = {Molecular beam scattering from solid surfaces},
	journal = {Progress in Surface Science},
	volume = {5},
	pages = {377-439},
	year = {1975},
	issn = {0079-6816},
	doi = {https://doi.org/10.1016/0079-6816(75)90001-5},
	url = {https://www.sciencedirect.com/science/article/pii/0079681675900015},
	author = {S.L. Bernasek and G.A. Somorjai},
}

@Article{ClausingZfP30,
	author={Clausing, P.},
	title={{\"U}ber die Strahlformung bei der Molekularstr{\"o}mung},
	journal={Zeitschrift f{\"u}r Physik},
	year={1930},
	month={Jul},
	day={01},
	volume={66},
	number={7},
	pages={471-476},
	issn={0044-3328},
	doi={10.1007/BF01402029},
	url={https://doi.org/10.1007/BF01402029},
}

@article{ClausingAdP32,
	author = {Clausing, P.},
	title = "{The Flow of Highly Rarefied Gases through Tubes of Arbitrary Length}",
	journal = {Journal of Vacuum Science and Technology},
	volume = {8},
	number = {5},
	pages = {636-646},
	year = {1971},
	month = {09},
	issn = {0022-5355},
	doi = {10.1116/1.1316379},
	url = {https://doi.org/10.1116/1.1316379},
}

@article{DaviesJPD83,
	doi = {10.1088/0022-3727/16/1/004},
	url = {https://dx.doi.org/10.1088/0022-3727/16/1/004},
	year = {1983},
	month = {jan},
	publisher = {},
	volume = {16},
	number = {1},
	pages = {1},
	author = {C M Davies and  C B Lucas},
	title = {The failure of theory to predict the density distribution of gas flowing through a tube under free molecular conditions},
	journal = {Journal of Physics D: Applied Physics},
}

@article{DavisJAP64,
	author = {Davis, Donald H. and Levenson, Leonard L. and Milleron, Norman},
	title = "{Effect of ``Rougher‐than‐Rough'' Surfaces on Molecular Flow through Short Ducts}",
	journal = {Journal of Applied Physics},
	volume = {35},
	number = {3},
	pages = {529-532},
	year = {1964},
	month = {03},
	doi = {10.1063/1.1713407},
	url = {https://doi.org/10.1063/1.1713407},
}

@techreport{deLeeuwUTN66,
	title={Density distribution of a molecular flux from a short cylindrical tube},
	author={De Leeuw, Jacob Henri and Gadamer, Ernst Oscar},
	year={1967},
	institution={Citeseer}
}

@inproceedings{Drullinger85,
	title={A recirculating oven for atomic beam frequency standards},
	author={Drullinger, RE and Glaze, DJ and Sullivan, DB},
	booktitle={39th Annual Symposium on Frequency Control},
	pages={13--17},
	year={1985},
	organization={IEEE},
	doi = {10.1109/FREQ.1985.200811},
	url = {https://doi.org/10.1109/FREQ.1985.200811},
}

@article{EdmondsJVST65,
	author = {Edmonds, T. and Hobson, J. P.},
	title = "{A Study of Thermal Transpiration Using Ultrahigh-Vacuum Techniques}",
	journal = {Journal of Vacuum Science and Technology},
	volume = {2},
	number = {4},
	pages = {182-197},
	year = {1965},
	month = {07},
	doi = {10.1116/1.1492423},
	url = {https://doi.org/10.1116/1.1492423},
}

@techreport{demarcus1956,
  title={THE PROBLEM OF KNUDSEN FLOW},
  author={DeMarcus, WC},
  year={1956},
  institution={Oak Ridge Gaseous Diffusion Plant, Tenn.}
}

@article{nusinzon1977gas,
  title={Gas flow in cylindrical capillaries under intermediate conditions},
  author={Nusinzon, LM and Porodnov, BT and Suetin, PE},
  journal={Fluid Dynamics},
  volume={12},
  number={1},
  pages={161--164},
  year={1977},
  publisher={Springer}
}

@article{Smoluchowski1910kinetischen,
  title={Zur kinetischen theorie der transpiration und diffusion verd{\"u}nnter gase},
  author={v. Smoluchowski, M},
  journal={Annalen der Physik},
  volume={338},
  number={16},
  pages={1559--1570},
  year={1910},
  publisher={WILEY-VCH Verlag Leipzig}
}

@article{EiblJVST98,
	author = {Eibl, C. and Lackner, G. and Winkler, A.},
	title = "{Quantitative characterization of a highly effective atomic hydrogen doser}",
	journal = {Journal of Vacuum Science and Technology A},
	volume = {16},
	number = {5},
	pages = {2979-2989},
	year = {1998},
	month = {09},
	issn = {0734-2101},
	doi = {10.1116/1.581449},
	url = {https://doi.org/10.1116/1.581449},
}

@article{FloryJAP93,
	author = {Flory, C. A. and Cutler, L. S.},
	title = "{Integral equation solution of low‐pressure transport of gases in capillary tubes}",
	journal = {Journal of Applied Physics},
	volume = {73},
	number = {4},
	pages = {1561-1569},
	year = {1993},
	month = {02},
	doi = {10.1063/1.353236},
	url = {https://doi.org/10.1063/1.353236},
}

@article{GiordmaineJAP60,
	author = {Giordmaine, J. A. and Wang, T. C.},
	title = "{Molecular Beam Formation by Long Parallel Tubes}",
	journal = {Journal of Applied Physics},
	volume = {31},
	number = {3},
	pages = {463-471},
	year = {1960},
	month = {3},
	issn = {0021-8979},
	doi = {10.1063/1.1735609},
	url = {https://doi.org/10.1063/1.1735609},
}

@article{HahnRSI22,
	author = {Hahn, Raphaël and Battard, Thomas and Boucher, Oscar and Picard, Yan J. and Lignier, Hans and Comparat, Daniel and Keriel, Nolwenn-Amandine and Lopez, Colin and Oswald, Emanuel and Reveillard, Morgan and Viteau, Matthieu},
	title = "{Comparative analysis of recirculating and collimating cesium ovens}",
	journal = {Review of Scientific Instruments},
	volume = {93},
	number = {4},
	pages = {043302},
	year = {2022},
	month = {04},
	issn = {0034-6748},
	doi = {10.1063/5.0085838},
	url = {https://doi.org/10.1063/5.0085838},
}

@article{HanesJAP60,
	author = {Hanes, G. R.},
	title = "{Multiple Tube Collimator for Gas Beams}",
	journal = {Journal of Applied Physics},
	volume = {31},
	number = {12},
	pages = {2171-2175},
	year = {1960},
	month = {12},
	doi = {10.1063/1.1735519},
	url = {https://doi.org/10.1063/1.1735519},
}

@article{HelmerJVST67II,
	author = {Helmer, John C.},
	title = "{Solution of Clausing’s Integral Equation for Molecular Flow}",
	journal = {Journal of Vacuum Science and Technology},
	volume = {4},
	number = {6},
	pages = {360-363},
	year = {1967},
	month = {11},
	issn = {0022-5355},
	doi = {10.1116/1.1492563},
	url = {https://doi.org/10.1116/1.1492563},
}

@article{HobsonJVST69,
	author = {Hobson, J. P.},
	title = "{Surface Smoothness in Thermal Transpiration at Very Low Pressures}",
	journal = {Journal of Vacuum Science and Technology},
	volume = {6},
	number = {1},
	pages = {257-259},
	year = {1969},
	month = {01},
	doi = {10.1116/1.1492674},
	url = {https://doi.org/10.1116/1.1492674},
}

@incollection{HurlbutRRMB59,
	title={Molecular scattering at the solid surface},
	author={Hurlbut, FC},
	booktitle={Recent Research in Molecular Beams},
	pages={145--156},
	year={1959},
	publisher={Academic Press}
}

@article{KnudsenAdP09I,
	author = {Knudsen, Martin},
	title = {Die Gesetze der Molekularströmung und der inneren Reibungsströmung der Gase durch Röhren},
	journal = {Annalen der Physik},
	volume = {333},
	number = {1},
	pages = {75-130},
	doi = {https://doi.org/10.1002/andp.19093330106},
	url = {https://onlinelibrary.wiley.com/doi/abs/10.1002/andp.19093330106},
	year = {1909},
}

@article{KnudsenAdP09II,
	author = {Knudsen, Martin},
	title = {Die Molekularströmung der Gase durch Offnungen und die Effusion},
	journal = {Annalen der Physik},
	volume = {333},
	number = {5},
	pages = {999-1016},
	doi = {https://doi.org/10.1002/andp.19093330505},
	url = {https://onlinelibrary.wiley.com/doi/abs/10.1002/andp.19093330505},
	year = {1909},
}

@article{KurepaJAP81,
	author = {Kurepa, M. V. and Lucas, C. B.},
	title = "{The density gradient of molecules flowing along a tube}",
	journal = {Journal of Applied Physics},
	volume = {52},
	number = {2},
	pages = {664-669},
	year = {1981},
	month = {02},
	issn = {0021-8979},
	doi = {10.1063/1.328844},
	url = {https://doi.org/10.1063/1.328844}
}

@article{LevdanskyIJHMT08,
	title = {Effect of surface diffusion on transfer processes in heterogeneous systems},
	journal = {International Journal of Heat and Mass Transfer},
	volume = {51},
	number = {9},
	pages = {2471-2481},
	year = {2008},
	issn = {0017-9310},
	doi = {https://doi.org/10.1016/j.ijheatmasstransfer.2007.08.009},
	url = {https://www.sciencedirect.com/science/article/pii/S0017931007005303},
	author = {V.V. Levdansky and J. Smolik and P. Moravec},
}

@article{LibudaRSI00,
	author = {Libuda, J. and Meusel, I. and Hartmann, J. and Freund, H.-J.},
	title = "{A molecular beam/surface spectroscopy apparatus for the study of reactions on complex model catalysts}",
	journal = {Review of Scientific Instruments},
	volume = {71},
	number = {12},
	pages = {4395-4408},
	year = {2000},
	month = {12},
	issn = {0034-6748},
	doi = {10.1063/1.1318919},
	url = {https://doi.org/10.1063/1.1318919},
}

@article{LoganJCP66,
	author = {Logan, R. M. and Stickney, R. E.},
	title = "{Simple Classical Model for the Scattering of Gas Atoms from a Solid Surface}",
	journal = {The Journal of Chemical Physics},
	volume = {44},
	number = {1},
	pages = {195-201},
	year = {1966},
	month = {01},
	issn = {0021-9606},
	doi = {10.1063/1.1726446},
	url = {https://doi.org/10.1063/1.1726446},
}

@book{Lucas2013,
	title={Atomic and molecular beams: production and collimation},
	author={Lucas, Cyril Bernard},
	year={2013},
	publisher={CRC press}
}

@article{LucasVacuum73,
	title = {The production of intense atomic beams},
	journal = {Vacuum},
	volume = {23},
	number = {11},
	pages = {395-402},
	year = {1973},
	issn = {0042-207X},
	doi = {https://doi.org/10.1016/0042-207X(73)92529-3},
	url = {https://www.sciencedirect.com/science/article/pii/0042207X73925293},
	author = {CB Lucas},
}

@article{MurphyJVST89,
	author = {Murphy, Daniel M.},
	title = "{Wall collisions, angular flux, and pumping requirements in molecular flow through tubes and microchannel arrays}",
	journal = {Journal of Vacuum Science and Technology A},
	volume = {7},
	number = {5},
	pages = {3075-3091},
	year = {1989},
	month = {09},
	issn = {0734-2101},
	doi = {10.1116/1.576317},
	url = {https://doi.org/10.1116/1.576317}
}

@article{OlanderJAP69I,
	author = {Jones, R. H. and Olander, D. R. and Kruger, V. R.},
	title = "{Molecular‐Beam Sources Fabricated from Multichannel Arrays. I. Angular Distributions and Peaking Factors}",
	journal = {Journal of Applied Physics},
	volume = {40},
	number = {11},
	pages = {4641-4649},
	year = {1969},
	month = {10},
	issn = {0021-8979},
	doi = {10.1063/1.1657245},
	url = {https://doi.org/10.1063/1.1657245},
}

@article{OlanderJAP69II,
	author = {Olander, Donald R.},
	title = "{Molecular‐Beam Sources Fabricated from Multichannel Arrays. II. Effect of Source Size and Alignment}",
	journal = {Journal of Applied Physics},
	volume = {40},
	number = {11},
	pages = {4650-4657},
	year = {1969},
	month = {10},
	issn = {0021-8979},
	doi = {10.1063/1.1657246},
	url = {https://doi.org/10.1063/1.1657246},
}

@article{OlanderJAP70III,
	author = {Olander, Donald R. and Kruger, Valerie},
	title = "{Molecular Beam Sources Fabricated from Multichannel Arrays. III. The Exit Density Problem}",
	journal = {Journal of Applied Physics},
	volume = {41},
	number = {7},
	pages = {2769-2776},
	year = {1970},
	month = {06},
	issn = {0021-8979},
	doi = {10.1063/1.1659313},
	url = {https://doi.org/10.1063/1.1659313},
}

@article{OlanderJAP70IV,
	author = {Olander, D. R. and Jones, R. H. and Siekhaus, W. J.},
	title = "{Molecular Beam Sources Fabricated from Multichannel Arrays. IV. Speed Distribution in the Centerline Beam}",
	journal = {Journal of Applied Physics},
	volume = {41},
	number = {11},
	pages = {4388-4391},
	year = {1970},
	month = {10},
	issn = {0021-8979},
	doi = {10.1063/1.1658472},
	url = {https://doi.org/10.1063/1.1658472},
}

@article{gray1992design,
  title={Design considerations for high-flux collisionally opaque molecular beams},
  author={Gray, David C and Sawin, Herbert H},
  journal={Journal of Vacuum Science \& Technology A: Vacuum, Surfaces, and Films},
  volume={10},
  number={5},
  pages={3229--3238},
  year={1992},
  publisher={American Vacuum Society}
}

@article{pan2013modeling,
  title={Modeling the resupply, diffusion, and evaporation of cesium on the surface of controlled porosity dispenser photocathodes},
  author={Pan, Zhigang and Jensen, Kevin L and Montgomery, Eric J},
  journal={Journal of Applied Physics},
  volume={114},
  number={10},
  year={2013},
  publisher={AIP Publishing}
}

@article{OlanderJAP70V,
	author = {Siekhaus, W. J. and Jones, R. H. and Olander, D. R.},
	title = "{Molecular Beam Sources Fabricated from Multichannel Arrays. V. Measurement of the Speed Distribution}",
	journal = {Journal of Applied Physics},
	volume = {41},
	number = {11},
	pages = {4392-4403},
	year = {1970},
	month = {10},
	issn = {0021-8979},
	doi = {10.1063/1.1658473},
	url = {https://doi.org/10.1063/1.1658473},
}

@book{Ramsey1956,
	title={Molecular beams},
	author={Ramsey, Norman},
	volume={20},
	year={1956},
	publisher={Oxford University Press}
}

@article{SchwarzSelingerJVST00,
	author = {Schwarz-Selinger, T. and von Keudell, A. and Jacob, W.},
	title = "{Novel method for absolute quantification of the flux and angular distribution of a radical source for atomic hydrogen}",
	journal = {Journal of Vacuum Science and Technology A},
	volume = {18},
	number = {3},
	pages = {995-1001},
	year = {2000},
	month = {05},
	issn = {0734-2101},
	doi = {10.1116/1.582289},
	url = {https://doi.org/10.1116/1.582289},
}

@article{SteckelmacherJJAP74,
	doi = {10.7567/JJAPS.2S1.117},
	url = {https://dx.doi.org/10.7567/JJAPS.2S1.117},
	year = {1974},
	month = {jan},
	publisher = {},
	volume = {13},
	number = {S1},
	pages = {117},
	author = {W. Steckelmacher},
	title = {The Flow of Rarefied Gases in Vacum Systems and Problems of Standardization of Measuring Techniques},
	journal = {Japanese Journal of Applied Physics}
}

@article{ZugenmaierZaP66,
	title={Zur theorie der molekularstrahlerzeugung mit Hilfe zylindrischer Rohre},
	author={Zugenmaier, P},
	journal={Zeitschrift fur angewandte Physik},
	volume={20},
	number={3},
	pages={184},
	year={1966}
}

@article{matsushima2003angle,
  title={Angle-resolved measurements of product desorption and reaction dynamics on individual sites},
  author={Matsushima, Tatsuo},
  journal={Surface science reports},
  volume={52},
  number={1-2},
  pages={1--62},
  year={2003},
  publisher={Elsevier}
}

@article{king1975thermal,
  title={Thermal desorption from metal surfaces: A review},
  author={King, David A},
  journal={Surface Science},
  volume={47},
  number={1},
  pages={384--402},
  year={1975},
  publisher={Elsevier}
}

@TECHREPORT{1974mbld.rept....1A,
  author = {{Anderson}, J.~B.},
  title = {{Molecular beams from nozzle sources}},
  year = {1974},
  adsurl = {http://adsabs.harvard.edu/abs/1974mbld.rept....1A},
  institution = {NASA},
  booktitle = {Molecular beams and low density gas dynamics. (A75-28372 12-72) New
	York, Marcel Dekker, Inc., 1974, p. 1-91.},
  pages = {1-91}
}

@ARTICLE{steckelmacher1986knudsen,
  author = {Steckelmacher, W},
  title = {Knudsen flow 75 years on: the current state of the art for flow of
	rarefied gases in tubes and systems},
  journal = {Reports on Progress in Physics},
  year = {1986},
  volume = {49},
  pages = {1083},
  number = {10},
  publisher = {IOP Publishing}
}

@article{kersevan2019recent,
  title={Recent developments of Monte-Carlo codes molflow+ and synrad+},
  author={Kersevan, Roberto and Ady, Marton and others},
  journal={Proc. IPAC’19},
  pages={1327--1330},
  year={2019}
}

@article{linke2025concurrent,
  title={A Concurrent Multiscale Framework Coupling Direct Simulation Monte Carlo and Molecular Dynamics},
  author={Linke, Tim and Sterbentz, Dane and Gr{\o}nbech-Jensen, Niels and Delplanque, Jean-Pierre and Belof, Jonathan},
  journal={arXiv preprint arXiv:2506.01924},
  year={2025}
}

@BOOK{ScolesII,
      key          = {354685},
      editor       = {Scoles, Giacinto and Laine, Derek and Valbusa, Ugo},
      title        = {{A}tomic and molecular beam methods: {V}ol. 2},
      address      = {New York},
      publisher    = {Oxford University Press},
      reportid     = {PUBDB-2017-111702},
      isbn         = {0195042816},
      pages        = {XVII, 534 pages},
      year         = {1992},
      ddc          = {539.6028},
      shelfmark    = {P Sco 2},
      typ          = {PUB:(DE-HGF)3},
      url          = {https://bib-pubdb1.desy.de/record/354685},
}

@BOOK{ScolesI,
      key          = {336634},
      editor       = {Scoles, Giacinto and Bassi, Davide and Buck, Udo and Laine,
                      Derek},
      title        = {{A}tomic and molecular beam methods: {V}ol. 1},
      address      = {New York},
      publisher    = {Oxford Univ. Pr.},
      reportid     = {PUBDB-2017-111701},
      isbn         = {0195042808},
      pages        = {721 pages},
      year         = {1988},
      ddc          = {539.6028},
      shelfmark    = {P Sco 1},
      typ          = {PUB:(DE-HGF)3},
      url          = {https://bib-pubdb1.desy.de/record/336634},
}

@BOOK{pauly2013atom,
  title = {Atom, Molecule, and Cluster Beams II: Cluster Beams, Fast and Slow
	Beams, Accessory Equipment and Applications},
  publisher = {Springer Science \& Business Media},
  year = {2013},
  author = {Pauly, Hans},
  volume = {32}
}

@BOOK{Pauly,
  title = {{Atom, Molecule and Clusterbeams I: Basic Theory, Production and
	Detection of Thermal Beams}},
  publisher = {Springer-Verlag, Berlin},
  year = {2000},
  author = {Pauly, Hans}
}

@ARTICLE{2013RScI...84j6113J,
  author = {{Jana}, B. and {Majumder}, A. and {Thakur}, K.~B. and {Das}, A.~K.
	},
  title = {{Design principles of a linear array multi-channel effusive
	metal-vapor atom source}},
  journal = {Review of Scientific Instruments},
  year = {2013},
  volume = {84},
  pages = {106113-106113},
  number = {10},
  adsurl = {http://adsabs.harvard.edu/abs/2013RScI...84j6113J},
  doi = {10.1063/1.4825343}
}

@article{buckman1993spatial,
  title={Spatial profiles of effusive molecular beams and their dependence on gas species},
  author={Buckman, Stephen J and Gulley, RJ and Moghbelalhossein, M and Bennett, SJ},
  journal={Measurement Science and Technology},
  volume={4},
  number={10},
  pages={1143},
  year={1993},
  publisher={IOP Publishing}
}

@ARTICLE{2015RScI...86b3105S,
  author = {{Senaratne}, R. and {Rajagopal}, S.~V. and {Geiger}, Z.~A. and {Fujiwara},
	K.~M. and {Lebedev}, V. and {Weld}, D.~M.},
  title = {{Effusive atomic oven nozzle design using an aligned microcapillary
	array}},
  journal = {Review of Scientific Instruments},
  year = {2015},
  volume = {86},
  pages = {023105},
  number = {2},
  month = feb,
  eid = {023105},
  adsurl = {http://adsabs.harvard.edu/abs/2015RScI...86b3105S},
  archiveprefix = {arXiv},
  doi = {10.1063/1.4907401},
  eprint = {1407.6391},
  primaryclass = {physics.atom-ph}
}

@article{dotti2025robust,
  title={Robust high-temperature atomic beam source with a microcapillary array},
  author={Dotti, Peter and Chai, Xiao and Tanlimco, Jeremy L and Simmons, Ethan Q and Weld, David M},
  journal={Review of Scientific Instruments},
  volume={96},
  number={6},
  year={2025},
  publisher={AIP Publishing}
}

@article{li2013modified,
  title={A modified numerical method for the accurate calculation of molecular flow transmission probabilities and density distributions of cylindrical tubes},
  author={Li, Yanwu and Chen, Xuekang and Bai, Xiaohang and Che, Qinglun and Li, Yajuan},
  journal={Vacuum},
  volume={97},
  pages={60--64},
  year={2013},
  publisher={Elsevier}
}

@article{drullinger1991nist,
  title={The NIST optically pumped cesium frequency standard},
  author={Drullinger, Robert E and Glaze, David J and Lowe, JL and Shirley, Jon H},
  journal={IEEE Transactions on Instrumentation and Measurement},
  volume={40},
  number={2},
  pages={162--164},
  year={1991}
}

@ARTICLE{ross95,
  author = {Ross, K. J. and Sonntag, B.},
  title = {High temperature metal atom beam sources},
  journal = {Review of Scientific Instruments},
  year = {1995},
  volume = {66},
  pages = {4409-4433},
  number = {9},
  doi = {http://dx.doi.org/10.1063/1.1145337},
  url = {http://scitation.aip.org/content/aip/journal/rsi/66/9/10.1063/1.1145337}
}

@inproceedings{shiwei2008computer,
  title={Computer simulation of positional beaming effect of molecular flow in straight cylindrical pipeline by Monte Carlo method},
  author={Shiwei, Zhang and Jin, Han and Zhijun, Zhang},
  booktitle={2008 International Conference on Computer Science and Information Technology},
  pages={486--491},
  year={2008},
  organization={IEEE}
}

@article{he2025spatial,
  title={Spatial distribution of gas flux in molecular flow regime of linear differential pumping systems},
  author={He, Ziyang and Fan, Haitao and Bi, Hailin and Xu, Minggao and Wang, Xudi and Yang, Jiuzhong},
  journal={Vacuum},
  pages={114740},
  year={2025},
  publisher={Elsevier}
}

@book{laurendeau2005statistical,
  title={Statistical thermodynamics: fundamentals and applications},
  author={Laurendeau, Normand M},
  year={2005},
  publisher={Cambridge University Press}
}

@book{karniadakis2005microflows,
  title={Microflows and nanoflows: fundamentals and simulation},
  author={Karniadakis, George and Beskok, Ali and Aluru, Narayan},
  year={2005},
  publisher={Springer}
}

@article{krasuski1990gas,
  title={Gas flux distributions from cylindrical tubes in molecular flow},
  author={Krasuski, PT},
  journal={Vacuum},
  volume={41},
  number={7-9},
  pages={2129--2131},
  year={1990},
  publisher={Elsevier}
}

@article{li2019cascaded,
  title={Cascaded collimator for atomic beams traveling in planar silicon devices},
  author={Li, Chao and Chai, Xiao and Wei, Bochao and Yang, Jeremy and Daruwalla, Anosh and Ayazi, Farrokh and Raman, Chandra},
  journal={Nature communications},
  volume={10},
  number={1},
  pages={1831},
  year={2019},
  publisher={Nature Publishing Group UK London}
}

@article{zhang2025foam,
  title={Foam copper adsorber facilitates the collimation of rubidium atomic beam},
  author={Zhang, Gehui and Zhao, Bingquan and Lu, Xiangxiang and Zhao, Qianyun and Wang, Jieying and Liu, Jian and Liu, Weiren and Wei, Junxin and Wang, Dianliang and Chen, Shanshan},
  journal={Review of Scientific Instruments},
  volume={96},
  number={4},
  year={2025},
  publisher={AIP Publishing}
}

@book{Bernstein1982,
  editor    = {Bernstein, Richard B.},
  title     = {Atom--Molecule Collision Theory: A Guide for the Experimentalist},
  publisher = {Plenum Press},
  address   = {New York},
  year      = {1979}
}

@book{LevineBernstein1987,
  author    = {Levine, Raphael D. and Bernstein, Richard B.},
  title     = {Molecular Reaction Dynamics and Chemical Reactivity},
  publisher = {Oxford University Press},
  address   = {New York},
  year      = {1987}
}

@book{GoodmanWachman1976,
  author    = {Goodman, Frederick O. and Wachman, Henry Y.},
  title     = {Dynamics of Gas-Surface Scattering},
  publisher = {Academic Press},
  address   = {New York},
  year      = {1976}
}

@book{campargue2001atomic,
  editor = {Campargue, Roger},
  title = {Atomic and Molecular Beams: The State of the Art 2000},
  year = {2001},
  publisher = {Springer},
  isbn = {978-3-540-67378-1},
  doi = {10.1007/978-3-642-56800-8},
  address = {Berlin, Heidelberg},
  edition = {1st},
  url = {https://link.springer.com/book/10.1007/978-3-642-56800-8}
}

@article{hopkins2016versatile,
  title={A versatile dual-species Zeeman slower for caesium and ytterbium},
  author={Hopkins, SA and Butler, K and Guttridge, A and Kemp, S and Freytag, R and Hinds, EA and Tarbutt, MR and Cornish, SL},
  journal={Review of Scientific Instruments},
  volume={87},
  number={4},
  pages={043109},
  year={2016},
  publisher={AIP Publishing}
}

@article{kemp2016production,
  title={Production and characterization of a dual species magneto-optical trap of cesium and ytterbium},
  author={Kemp, SL and Butler, KL and Freytag, R and Hopkins, SA and Hinds, EA and Tarbutt, MR and Cornish, SL},
  journal={Review of Scientific Instruments},
  volume={87},
  number={2},
  pages={023105},
  year={2016},
  publisher={AIP Publishing}
}

@article{guevremont2000design,
  title={Design and characterization of collimated effusive gas beam sources: Effect of source dimensions and backing pressure on total flow and beam profile},
  author={Guevremont, Jeffrey M and Sheldon, Stanley and Zaera, Francisco},
  journal={Review of Scientific Instruments},
  volume={71},
  number={10},
  pages={3869--3881},
  year={2000},
  publisher={AIP}
}

@article{majumder2009generation,
  title={Generation of a long wedge-shaped barium atomic beam and its density characterization},
  author={Majumder, A and Jana, B and Kathar, PT and Das, AK and Mago, VK},
  journal={Vacuum},
  volume={83},
  number={6},
  pages={989--995},
  year={2009},
  publisher={Elsevier}
}

@INBOOK{Brau02,
  chapter = {What Brightness means},
  pages = {20},
  title = {The Physics and Applications of High Brightness Electron Beam},
  publisher = {Proceedings of the ICFA Workshop, Chia Laguna, Sardinia},
  year = {2002},
  editor = {World Scientific},
  author = {C. A. Brau}
}

@article{rhee1992refined,
  title={Refined definition of the beam brightness},
  author={Rhee, MJ},
  journal={Physics of Fluids B: Plasma Physics},
  volume={4},
  number={6},
  pages={1674--1676},
  year={1992},
  publisher={American Institute of Physics}
}

@ARTICLE{LEJEUNEC1980,
  author={Lejeune C.; Aubert J.},
  title={EMITTANCE AND BRIGHTNESS: DEFINITIONS AND MEASUREMENTS},
  journal={Adv. Electron. Electron Phys. Suppl.},
  volume= {13},
  pages={159},
  year={1980},
  affiliation={UNIV. PARIS, INST. ELECTRON. FONDAMENTALE/ORSAY/FRA},
  descriptors={FAISCEAU PARTICULE CHARGEE; PROPRIETE OPTIQUE; EMITTANCE; BRILLANCE; MESURE; ELECTRONIQUE},
  subject={Electronique[001D03]},
  language={English},
  document_type={Article},
  inist_number={PASCAL8130103080} 
  }

@article{de2021spider,
  title={SPIDER Cs Ovens functional tests},
  author={De Muri, M and Rizzolo, A and Sartori, E and Cristofaro, S and Barbisan, M and Fadone, M and Ravarotto, D and Rizzieri, R and Capobianco, R and Cinetto, P and others},
  journal={Fusion Engineering and Design},
  volume={167},
  pages={112331},
  year={2021},
  publisher={Elsevier}
}

@article{seccombe2001design,
  title={The design and performance of an effusive gas source of conical geometry for photoionization studies},
  author={Seccombe, DP and Collins, SA and Reddish, TJ},
  journal={Review of Scientific Instruments},
  volume={72},
  number={6},
  pages={2550--2557},
  year={2001},
  publisher={AIP}
}

@FOOTNOTE{Note1,key="Note1",note="There is apparently a typo in Eq.~(31) of Ref.~\protect \rev@citealpnum  {OlanderJAP70III}, in the last term of the equation, where the prefactor $\left [\left (1-\zeta _1\right )/\zeta _0\right ]$ should be replaced by $\left (1-\zeta _1\right )$. This error was also copied in Ref.~\protect \rev@citealpnum  {Lucas2013} (Eq.~(8.39) p. 213)."}

@FOOTNOTE{Note2,key="Note2",note="In Ref.~\protect \rev@citealpnum  {Lucas2013}, there is apparently an unfortunate typo in the presentation of the angular profile (Eqs.~(8.33) and~(8.34), page 207), with a factor $\left (1-\alpha \right )$ instead of $\alpha $."}

@FOOTNOTE{Note3,key="Note3",note="Ref.~\protect \rev@citealpnum  {Lucas2013}, indicate that Zugenmaier’s own equation for H (his equation 35) is incorrect."}

\end{document}